\documentclass[3p,times,number,sort&compress]{elsarticle}

\journal{Annals of Physics}

\usepackage{amssymb}
\usepackage{amsthm}
\usepackage{amsmath}
\usepackage{bbm}

\newcommand{\av}[1]{\left\langle #1 \right\rangle} 

\def\trlog{trace-log}

\DeclareMathOperator{\sign}{sign}

\newlength{\graphwid} 
\def\showgraph#1#2{
\settowidth{\graphwid}{\includegraphics[#1,clip=true]{#2}}
\parbox[c]{\graphwid}{\includegraphics[#1,clip=true]{#2}}}

\usepackage[figuresright]{rotating}

\begin{document}

\begin{frontmatter}




\title{Majorana representation for dissipative spin systems}


\author[kit1]{P. Schad}
\author[landau,mfti]{Yu. Makhlin}
\author[kit1,mifi]{B.N. Narozhny}
\author[kit2,kit3]{G. Sch\"on}
\author[kit1]{A. Shnirman}

\address[kit1]{Institut f\"ur Theorie der Kondensierten Materie,
  Karlsruhe Institute of Technology, D-76131 Karlsruhe, Germany}

\address[landau]{Landau Institute for Theoretical Physics, acad. Semyonov 
av., 1a, 142432, Chernogolovka, Russia}

\address[mfti]{Moscow Institute of Physics and Technology, 141700, Dolgoprudny, Russia}

\address[mifi]{National Research Nuclear University MEPhI (Moscow
  Engineering Physics Institute), Kashirskoe shosse 31, 115409 Moscow,
  Russia}

\address[kit2]{Institut f\"ur Theoretische Festk\"orperphysik,
  Karlsruhe Institute of Technology, D-76131 Karlsruhe, Germany}

\address[kit3]{Institut f\"ur Nanotechnologie, Karlsruhe Institute of
  Technology, D-76021 Karlsruhe, Germany}

\begin{abstract}
  The Majorana representation of spin operators allows for efficient
  field-theoretical description of spin-spin correlation
  functions. Any N-point spin correlation function is equivalent to a
  2N-point correlator of Majorana fermions. For a certain class of
  N-point spin correlation functions (including ``auto'' and
  ``pair-wise'' correlations) a further simplification is possible, as
  they can be reduced to N-point Majorana correlators. As a specific
  example we study the Bose-Kondo model. We develop a path-integral
  technique and obtain the spin relaxation rate from a saddle point
  solution of the theory. Furthermore, we show that fluctuations
  around the saddle point do not affect the correlation functions as
  long as the latter involve only a single spin projection. For
  illustration we calculate the 4-point spin correlation function
  corresponding to the noise of susceptibility.
\end{abstract}


\end{frontmatter}



\section*{} 

Spin systems are notoriously difficult to describe using
field-theoretic methods due to non-Abelian nature of the spin
operators~\cite{altland}. Often one tries to circumvent the problem by
mapping the spins onto a system of either bosons or fermions, for
which a standard field theory can be developed~\cite{Tsvelik}. There
is no unique recipe for such an approach. Several formulations have
been put forth for solving specific problems, including the
Jordan-Wigner~\cite{JordanWigner} and
Holstein-Primakoff~\cite{Holstein} transformations, the
Martin~\cite{Martin} Majorana-fermion and Abrikosov~\cite{Avella12}
fermion representations, as well as the
Schwinger-boson~\cite{schwinger,arovas,read,wang} and
slave-fermion~\cite{affleck,marston,andrei,dagotto,wen,nayak}
techniques.

The Jordan-Wigner transformation is the only ``exact'' mapping between
the spin-$1/2$ and fermion operators as it preserves not only the
operator algebra but also the dimensionality of the Hilbert
space. However, it is a non-local transformation specific to one
spatial dimension~\cite{Tsvelik} where it is often applied to
Bethe-Ansatz-solvable models or their variations (a generalization to
higher dimensions does exist~\cite{fradkin,huerta}, but it lacks the
simplicity of the original approach). All other mappings suffer from
the following two problems: (i) the Hilbert space of the fermionic or
bosonic operators (the ``target'' Hilbert space) is enlarged as
compared to the original spin Hilbert space, and (ii) the resulting
theory in the fermionic or bosonic representation needs to be treated
perturbatively, which often leads to complicated vertex structures
(see e.g.~\cite{ZawFaz}). The former issue may be resolved by
additional constraints~\cite{wang,nayak} or by projecting out
unphysical states~\cite{Avella12} at the expense of further
complications such as the appearance of non-Abelian gauge
fields~\cite{read,wang,nayak,andrei}.

The Majorana-fermion representation, suggested by
Martin~\cite{Martin}, offers a possibility to avoid both types of
problems mentioned above: (i) The target Hilbert space is indeed enlarged, but merely 
contains two (or more) copies of the original physical spin Hilbert space
~\cite{Spencer68,Shnirman03}. Matrix
elements of physical quantities between different copies of the
original Hilbert space vanish and thus the correlation functions may
be evaluated directly in the target Hilbert space (this fact is often not fully appreciated; 
below we justify and further illustrate this statement).
(ii) The Martin transformation~\cite{Martin} represents the spin
operators in terms of bilinear combinations of Majorana fermions. The
resulting Majorana theory appears to be interacting (similarly to the
situation with the Jordan-Wigner transformation~\cite{JordanWigner}
and other fermionic
representations~\cite{affleck,marston,andrei,dagotto,wen,nayak}) and
$N$-point spin correlation functions are equivalent to $2N$-fermion
correlators. However, a certain class of $N$-point spin correlation
functions can be reduced to $N$-point Majorana
correlators~\cite{Mao03,Shnirman03}. In this case, complicated vertex
structures do not arise. This significant simplification applies to
correlation functions which involve only a single spin
(``auto-correlation'' functions) or an even number of spin operators
for each spin (``pair-wise'' correlations). In the present paper we
generalize this approach to higher-order correlation functions and
develop a technique for practical calculations based on the Keldysh
formalism~\cite{Kamenev}.

We illustrate our general arguments by a specific example. For a model
of paramagnetic spins coupled to an Ohmic bath~\cite{dutta} we
determine the four-point correlation function describing the noise of
susceptibility~\cite{Schad14}. This quantity is closely related to the
recently measured inductance fluctuations in
SQUIDs~\cite{SendelbachMcDermott2}. In addition, it is a direct
measure of non-Gaussian fluctuations that are relevant also in other
physical contexts~\cite{weissman88,weissman93,Burnett14}. The noise of
susceptibility is distinct from the better-known four-point correlator
``second noise'', which always comprises a significant Gaussian
contribution~\cite{kogan,beckspruit,SeidlerSolin}.

The paper is divided into two major parts. In Section~\ref{sec:majrep}
we introduce the Majorana representation of spin operators and address
general issues related to this approach. In Subsection~\ref{theorem}
we discuss the fact that spin-spin correlation functions can be
directly calculated as correlations of the Majorana fermions despite
the enlargement of the fermion Hilbert space. In Subsection~\ref{auto}
we demonstrate the simplification of the theory for the case of auto-
and pair-wise-correlations. In Subsections~\ref{sec:keldysh}
and~\ref{mpath} we formulate of the above correlators within the
Keldysh path-integral approach.

In Section~\ref{sec:BoseKondo} we apply our general arguments to the
problem of the noise of higher-order spin correlators in the
Bose-Kondo model. Here we introduce the path-integral formalism in the
Matsubara representation and show that certain higher-order spin
correlators can be calculated in the saddle-point approximation.
Further technical details are provided in the Appendices.
In~\ref{pert}, we present the traditional diagrammatic perturbation
theory. In~\ref{sec:pathint_keldysh} we justify the saddle-point
approximation used in Section~\ref{sec:BoseKondo}. Finally,
in~\ref{sec:Anonzero} we analyze a remarkable gauge freedom in our
model.

\section{Majorana Representation for Spin Operators} 
\label{sec:majrep}
\subsection{The Martin transformation}

In this paper we focus on the following Majorana representation
of the spin-$1/2$ operators introduced by Martin in 1959~\cite{Martin}
in the framework of generalized classical dynamics:
\begin{equation} 
\label{Majrep}
\hat{S}^\alpha = -\frac{i}{2} \epsilon_{\alpha\beta\gamma} \hat\eta_\beta \hat\eta_\gamma\ , 
\qquad 
\hat{S}^x= -i\hat\eta_y \hat\eta_z\ ,
\quad 
\hat{S}^y= -i\hat\eta_z \hat\eta_x\ ,
\quad 
\hat{S}^z= -i\hat\eta_x \hat\eta_y\ .
\end{equation}
The Majorana operators obey the Clifford algebra
\begin{equation}
\{\hat\eta_\alpha, \hat\eta_\beta\}=\delta_{\alpha\beta}\ , \qquad\hat \eta_\alpha^2=1/2\ ,
\end{equation}
where $\{.,.\}$ 
denotes the anticommutator.  The Majorana representation has been used
in a variety of physical contexts,
cf.~Refs.~\cite{Berezin77,Tsvelik,Mao03,Shnirman03}.  (We remark that
another normalization of the Majorana operators, $\hat\eta^2=1$, used
in Ref.~\cite{Shnirman03} only changes some numerical prefactors at
intermediate stages of the calculation.)

The above representation with the real Majorana fermion operators,
\begin{equation}
\hat\eta^\dagger=\hat\eta,
\end{equation}
perfectly reproduces the SU$(2)$ algebra of the operators $\hat{S}^\alpha$ 
\begin{equation}
\left[\hat{S}^\alpha,\hat{S}^\beta\right]=i \epsilon_{\alpha\beta\gamma} \hat{S}^\gamma ,
\end{equation}
and explicitly preserves the spin-rotation symmetry. 

Applying standard field-theoretical methods to fermionic systems
implies the existence of a Fock space. In a faithful representation of
a spin $1/2$, the dimensionality of the fermionic Fock space should
coincide with the dimensionality of the Hilbert space of the spin. For
example, the Jordan-Wigner transformation represents a system of $N$
spins-$1/2$ with the $2^N$-dimensional Hilbert space in terms of $N$
fermions with the $2^N$-dimensional Fock space. However, each
Jordan-Wigner fermion may be expressed in terms of {\it two} Majorana
fermions. In contrast, the Martin transformation (\ref{Majrep})
represents each spin in terms of {\it three} Majorana fermions and
hence does not preserve the dimensionality of the spin Hilbert space.

One method of dealing with this issue is to express the Majorana (or
``real fermion'') $\hat\eta$-operators in terms of "complex" or "Dirac" fermions 
(this is a common slang used to distinguish usual fermions from Majoranas; 
these fermions do not necessarily obey the Dirac equation) and use
their respective standard Fock spaces. This requires an even number of
Majorana fermions. The simplest possibility is to add one auxiliary
Majorana (cf.~the drone-fermion representations of
Refs.~\cite{Martin,Kenan66,Doniach67,Spencer68}). In this case, the
resulting target Hilbert space is $4$-dimensional and can be
interpreted as two copies of the spin Hilbert
space~\cite{Spencer68,Shnirman03}. To make the spin-space isotropy
even more explicit, one could add (and pair) an extra Majorana to each
of the three $\hat\eta_\alpha$; this, however, results in an
$8$-dimensional Hilbert space, equivalent to four copies of
the spin Hilbert space. In the following subsection we demonstrate that regardless of the 
construction of the Hilbert space, spin correlation functions 
coincide with Majorana-fermion correlators.

\subsection{Equivalence of spin and Majorana correlation functions}
\label{theorem}

Now we show explicitly that correlation functions of spin-$1/2$
operators can be directly computed in the Majorana
representation~(\ref{Majrep}) (regardless of the explicit construction
of the Majorana Hilbert space). Indeed, the equation (\ref{Majrep})
gives a representation of SU(2) with ${\bf\hat{S}}^2=3/4$. Thus, it is
necessarily a direct sum of a certain number of irreducible,
two-dimensional spin-$1/2$ representations, i.e, we obtain an integer
number of copies of the spin (the actual number is determined by the
particular choice of the number of auxiliary Majorana operators). The
spin operators~(\ref{Majrep}) do not switch between the copies, thus
any trace of an operator built out of spin operators (\ref{Majrep}) is
given by the number of copies times the trace in a two-dimensional
spin space. This leads directly to the desired result as we discuss in
more detail below.

Let us remind the reader that any function of the spin-$1/2$ operators
(i.e., of Pauli matrices) is in fact a linear function. This follows
from the following relation for the spin-$1/2$ operators
\begin{equation}
\hat{S}^\alpha \hat{S}^\beta = \frac{i}{2}\epsilon_{\alpha\beta\gamma} \hat{S}^\gamma 
+\frac{1}{4}\delta_{\alpha\beta}.
\label{Sred}
\end{equation}
For an arbitrary number of operators describing the same spin this
implies that
\begin{equation}
\hat{S}^{\alpha_1}\ldots\hat{S}^{\alpha_n} = a_{\alpha} \hat{S}^{\alpha} + a_0, 
\qquad 
a_0,a_x,a_y,a_z \in {\mathbb C}\,,
\label{sprod}
\end{equation}
where the complex numbers $a_0$, $a_x$, $a_y$, and $a_z$ depend on the
sequence $\{\alpha_i\}$. Since spin operators are traceless, the
trace over the $2$-dimensional ($d_S=2$) spin Hilbert space
$\text{Tr}_S$ of the above combination of spin operators is given by
the constant term $a_0$:
\begin{equation}
\text{Tr}_\text{S}\left\{\hat{S}^{\alpha_1}\ldots\hat{S}^{\alpha_n}\right\} 
= d_S a_0(\alpha_1,...,\alpha_n), \qquad d_S=2.
\label{ts}
\end{equation}

The Martin representation (\ref{Majrep}) preserves the commutation
relations as well as the relation (\ref{Sred}), and hence the equality
(\ref{sprod}) remains valid in the Majorana representation (where the
spin operators should be understood as Majorana bilinears). In
particular, the coefficients $a_i$ remain the same. The trace over the
Majorana Hilbert space can be performed by noting that Majorana
bilinears are traceless due to their anticommutation properties. Again
[cf. Eq.~(\ref{ts})], the only remaining contribution is given by the
constant term $a_0$:
\begin{equation}
\text{Tr}_\text{M}\left\{\hat{S}^{\alpha_1}\ldots\hat{S}^{\alpha_n}\right\} 
= d_M a_0(\alpha_1,...,\alpha_n),
\label{tm}
\end{equation}
where $d_M$ is the dimension of the Majorana Hilbert space. Thus, {\it
  tracing an arbitrary product of the spin-$1/2$ operators over the
  spin and Majorana Hilbert spaces yields the same result up to a
  numerical factor, determined by the dimensionalities of the Hilbert
  spaces)}.

Similar arguments were put forth in Ref.~\cite{Spencer68} in the
context of the drone-fermion representation. The results of this
Subsection were implied in Refs.~\cite{Mao03,Shnirman03} and given without
proof in Ref.~\cite{Schad14}.

The above statement can be readily generalized to an arbitrary
ensemble of spins. Indeed, any function of spin operators is still
linear in the components of each spin. If the operators on the left-hand side of
Eq.~(\ref{sprod}) describe more than one spin, then on the right-hand
side additional terms appear, which contain all possible products of
operators related to different spins (for example, in the case of two
spins the right-hand side reads
$a_0+a_\alpha\hat{S}_1^\alpha+b_\alpha\hat{S}_2^\alpha+c_{\alpha\beta}\hat{S}_1^\alpha\hat{S}_2^\beta$).
However, all such additional terms are still traceless, and hence the
only change in Eqs.~(\ref{ts}) and (\ref{tm}) will be in the constants
$d_S$ and $d_M$.

Consider now a real-time spin auto-correlation function:
\begin{equation}
\label{scf0}
\av{\hat{S}^{\alpha_1}(t_1)\ldots \hat{S}^{\alpha_n}(t_n)}
\equiv
\frac{\text{Tr}\left\{\hat{S}^{\alpha_1}(t_1)\ldots\hat{S}^{\alpha_n}(t_n)\hat\rho\right\}}
     {\text{Tr}\left\{\hat\rho\right\}},
\end{equation}
where $\hat\rho=\exp(-\beta\hat{H})$ is the non-normalized Gibbs density
matrix. If in addition the spins are coupled to other degrees of freedom 
denoted collectively by $X$ (this is also described by $\hat{H}$), then following the 
line of arguments presented above we can write $\hat \rho=\hat\rho_0(X) + \hat{S}^{\alpha}\hat\rho_\alpha(X)$,
where $\hat\rho_0(X)$ and $\hat\rho_\alpha(X)$ are matrices in the $X$ space. Moreover,
$\hat\rho_0(X)$ is the reduced density matrix describing the rest of the system (the $X$-degrees of
freedom). In this case, the partition function
may be written as
\begin{equation}
\label{z1}
\text{Tr}\left\{\hat\rho\right\} 
= \text{Tr}_\text{X} \text{Tr}_\text{S} \left\{\hat\rho\right\} 
= \text{Tr}_\text{X} \text{Tr}_\text{S} \left\{\hat\rho_0(X) + \hat{S}^{\alpha}\hat\rho_\alpha(X)\right\} 
= d_s \text{Tr}_\text{X} \left\{\hat\rho_0(X)\right\} \ .
\end{equation}
Now, each Heisenberg spin operator $\hat{S}^{\alpha_i}(t_i)$ is related to
the Schr\"odinger operators (i.e., the Pauli matrices) by the time
evolution operator
$\hat{U}(t,t')=\exp[-i\int^t_{t'}d\tau\hat{H}(\tau)]$. The latter can
be expanded similarly to the density matrix. 
Thus we can write
$\hat{S}^{\alpha_1}(t_1)\ldots \hat{S}^{\alpha_n}(t_n)\hat \rho = \hat{A}_0(X)+\hat{S}^{\alpha}\hat{A}_\alpha(X)$, 
where $\hat{A}_0(X)$ and $\hat{A}_\alpha(X)$ are matrices in $X$-space.
As a result, we can
formally perform the trace over the spin variables in the
auto-correlation function (\ref{scf0}) and find
\begin{equation}
\label{scf1}
\av{\hat{S}^{\alpha_1}(t_1)\ldots \hat{S}^{\alpha_n}(t_n)}_\text{S}
= \frac{\text{Tr}_\text{X}\text{Tr}_\text{S}\left\{\hat{A}_0(X)+\hat{S}^{\alpha}\hat{A}_\alpha(X)\right\}}{d_S \text{Tr}_\text{X} \left\{\hat\rho_0(X)\right\}}
= \frac{\text{Tr}_\text{X}\left\{\hat{A}_0(X)\right\}}{\text{Tr}_\text{X}\left\{\hat\rho_0(X)\right\}}\  .
\end{equation}

In the Majorana representation the structure of the above equations
remains the same. The averages can be formally calculated similarly to
Eq.~(\ref{tm}). As a result, we arrive at the expression
\begin{equation}
\label{scf2}
\av{\hat{S}^{\alpha_1}(t_1)\ldots \hat{S}^{\alpha_n}(t_n)}_\text{M}
= \frac{\text{Tr}_\text{X}\text{Tr}_\text{M}\left\{\hat{A}_0(X)+\hat{S}^{\alpha}\hat{A}_\alpha(X)\right\}}{d_M \text{Tr}_\text{X}\left\{\hat\rho_0(X)\right\}}
= \frac{\text{Tr}_\text{X}\left\{\hat{A}_0(X)\right\}}{\text{Tr}_\text{X}\left\{\hat\rho_0(X)\right\}},
\end{equation}
which is identical to Eq.~(\ref{scf1}). The generalization to the case
of general (multi-spin) correlation function is straightforward (see above).
Thus we have demonstrated that spin correlation functions can be calculated with
the help of the Majorana representation (\ref{Majrep}) without any projection onto
"physical" states:
\begin{equation}
\label{theo}
\av{\hat{S}^{\alpha_1}(t_1)\ldots \hat{S}^{\alpha_n}(t_n)}_\text{S}
=
\av{\hat{S}^{\alpha_1}(t_1)\ldots \hat{S}^{\alpha_n}(t_n)}_\text{M}.
\end{equation}
The operators on the right-hand of Eq.~(\ref{theo}) represent the
Majorana bilinears~ (\ref{Majrep}).

\subsection{Simplified representation for autocorrelation functions}
\label{auto}

The arguments presented in the previous Section show that any
$N$-point spin correlation function may be represented in terms of a
$2N$-point correlation function of Majorana fermions. Here we
demonstrate that for a certain class of correlation functions this
correspondence can be significantly simplified
\cite{Mao03,Shnirman03,Coleman95}.

Let us rewrite Eq.~(\ref{Majrep}) as follows
\begin{equation}
\label{Majrep2}
\hat{S}^\alpha = \hat{\Theta} \hat{\eta}_\alpha, \qquad 
\hat{\Theta}=- 2i\hat{\eta}_x\hat{\eta}_y\hat{\eta}_z, \qquad
\hat{\Theta}^2=1/2.
\end{equation}
One can easily verify that the operator $\Theta$ commutes with
all three Majorana operators $\eta_\alpha$. Consequently, this
operator also commutes with any Hamiltonian expressed in terms of
$\eta_\alpha$, and thus the corresponding Heisenberg operator is
time-independent. The averaged product of a pair of spin operators can
now be represented as follows
\begin{equation}\label{eq:Trick}
\av{\hat{S}^\alpha(t))\hat{S}^\beta(t')}_\text{M}
= \av{\hat{\Theta}\hat{\eta}_\alpha(t) \hat{\Theta}\hat{\eta}_\beta(t')} 
= \frac 12 \av{\hat{\eta}_\alpha(t)\hat{\eta}_\beta(t')}.
\end{equation}
Thereby a two-point spin correlation function reduces to a {\it
  two}-point (rather than four-point) Majorana-fermion correlation.

Unfortunately, the above relation cannot be directly generalized to
time-ordered correlators (or Green's functions) due to the fact that
spin and Majorana operators are influenced by time ordering in different ways.
Explicitly, a time-ordered average of two spin operators is given by
\begin{equation}
\label{av1}
\av{T\left\{\hat{S}^\alpha(t)\hat{S}^\beta(t')\right\}}_\text{M} 
\equiv
\begin{cases}
\av{\hat{S}^\alpha(t)\hat{S}^\beta(t')}, & t>t' \cr
\av{\hat{S}^\beta(t')\hat{S}^\alpha(t)}, & t<t'
\end{cases}
=
\frac{1}{2}
\begin{cases}
\av{\hat{\eta}_\alpha(t)\hat{\eta}_\beta(t')}, & t>t' \cr
\av{\hat{\eta}_\beta(t')\hat{\eta}_\alpha(t)}, & t<t'
\end{cases},
\end{equation}
where the latter expression differs from the time-ordered average of
the two Majorana-fermion operators by the absence of the minus sign in the
lower line. 

While this problem can be circumvented by introducing Green's
functions of the operator $\hat{\Theta}$ \cite{Cabrera13}, we consider
here a simpler approach. The missing sign can be compensated for with
the help of an auxiliary Majorana fermion $\hat{m}$ that anti-commutes
with the three operators $\hat{\eta}_\alpha$, i.e.,
$\{\hat\eta,\hat{m}\}=0$ and $\hat{m}^2= 1/2$. Since any Hamiltonian
will be expressed in terms of bilinears of $\hat{\eta}_\alpha$, the
operator $\hat{m}$ commutes with the Hamiltonian and, thus, is time
independent.  We keep, however, its formal time argument in order to
be able to treat time-ordered operator products correctly. We notice
that
\[
\av{T\left\{im(t)\eta_\alpha(t) im(t')\eta_\beta(t')\right\}}
\equiv
\begin{cases}
\av{im(t)\eta_\alpha(t) im(t')\eta_\beta(t')}, & t>t' \cr
\av{im(t')\eta_\beta(t') im(t)\eta_\alpha(t)}, & t<t'
\end{cases}
=
\frac{1}{2}
\begin{cases}
\av{\eta_\alpha(t) \eta_\beta(t')}, & t>t' \cr
\av{\eta_\beta(t') \eta_\alpha(t)}, & t<t'
\end{cases}.
\]
Thus we arrive at the identity
\begin{equation}
\label{2CorId}
\av{T\left\{\hat{S}^\alpha(t) \hat{S}^\beta(t')\right\}}_\text{M}
=  \av{T\left\{i\hat{m}(t)\hat\eta_\alpha(t)\, i\hat{m}(t')\hat\eta_\beta(t')\right\}}.
\end{equation}
Here the $2$-point spin correlator is again expressed in terms of a
$4$-point Majorana correlation function. However, in contrast to the
direct application of the Martin transformation (\ref{Majrep}), here
two of the Majorana operators $\hat{m}(t)$ and $\hat{m}(t')$ do not
have any dynamical properties and only serve the purpose of writing
the time ordering (\ref{av1}) in a compact form.

Similarly, we can use the auxiliary operator $\hat{m}(t)$ to express
higher-order correlation functions. For a $4$-point spin correlator we
find
\begin{equation}
\label{4CorId}
\av{T\left\{\hat{S}^\alpha(t_1)\hat{S}^\beta(t_1')
            \hat{S}^\delta(t_2)\hat{S}^\gamma(t_2')\right\}}_\text{M} 
= 
\av{T\left\{\hat{m}(t_1)\hat\eta_\alpha(t_1)\hat{m}(t_1')\hat\eta_\beta(t_1') 
            \hat{m}(t_2)\hat\eta_\delta(t_2)\hat{m}(t_2')\hat\eta_\gamma(t_2')\right\}}.
\end{equation}

The correlation functions (\ref{2CorId}) and (\ref{4CorId}) are in
fact auto-correlation functions in the sense that they involve
operators describing the same spin. Clearly, the simplification
(\ref{eq:Trick}) cannot be extended to different spins since the
operators $\hat{\Theta}_1$ and $\hat{\Theta}_2$ anti-commute and
$\hat{\Theta}_1 \hat{\Theta}_2 \neq 1$. Therefore, at the level of
$2$-point correlation functions the simplification described in this
section applies to auto-correlation functions only. Generalizing this
technique to higher-order correlation functions, we can compute
autocorrelators, such as the $4$-point function (\ref{4CorId}), as
well as ``pair-wise'' correlators comprised of pairs of operators for
each spin, such as
\[
\av{T\left\{\hat{S}_1^\alpha(t_1)\hat{S}_2^\beta(t_1')
            \hat{S}_1^\delta(t_2)\hat{S}_2^\gamma(t_2')\right\}}_\text{M}.
\]
Other correlators, such as
$\av{T\left\{\hat{S}_1^\alpha(t)\hat{S}_2^\beta(t')\right\}}$, have to
be computed by different methods.

\subsection{Spin correlation functions in the Keldysh formalism} 
\label{sec:keldysh}

Real-time correlation functions at finite temperatures can be
conveniently computed within the Keldysh formalism
\cite{altland,Kamenev}. The calculation amounts to
finding the generating functional $Z_\lambda$~\cite{Kamenev} and then
taking the derivative with respect to the source fields.

For a spin system, the generating functional may be defined as follows
\begin{equation}
\label{zs}
Z_\lambda = \int D[\dots] 
\exp\left\{i{\cal S}_0+i\int\limits_C dt 
\left(\lambda_\alpha^{cl} S_\alpha^{q}+\lambda_\alpha^{q} S_\alpha^{cl}\right)\right\},
\end{equation}
where $D[\dots]$ denotes the appropriate measure of integration
whereas ${\cal S}_0$ represents the action of the model under
consideration.  In particular one could choose to integrate over the
SU$(2)$ group manifold and $D[\dots]$ would represent then the
appropriate Haar measure~\cite{altland}. In this paper we choose a
more straightforward method of integrating over real Grassmann
variables representing the Majorana operators of~(\ref{Majrep}). In
(\ref{zs}) $\lambda_\alpha^{cl(q)}$ are the source fields.  The
superscripts $cl$ and $q$ refer to the ``classical'' and ``quantum''
variables \cite{Kamenev} that are defined as the sum and difference of
the corresponding fields belonging to the upper ($u$) and lower ($d$)
branch of the Keldysh contour
\[
S_\alpha^{cl(q)}=\frac{1}{\sqrt{2}}\left(S_\alpha^{u}\pm S_\alpha^{d}\right),
\quad  
\lambda_\alpha^{cl(q)}=\frac{1}{\sqrt{2}}\left(\lambda_\alpha^{u}\pm\lambda_\alpha^{d}\right).
\]
The ''classical'' source term defined in this way describes the
physical probing field, $\lambda_\alpha^{cl}\equiv\sqrt{2}B_\alpha$,
while the ''quantum'' term is only needed to construct the correlation
function and is set to zero at the end of the calculation.

Taking the derivative of the functional (\ref{zs}) with respect to the
source fields $\lambda_\alpha^{cl(q)}$, one finds the spin correlation
functions. In particular, the one-point function defines the
magnetization
\begin{equation}
\label{mag}
\sqrt{2}{\cal M}^\alpha = \av{S^{\alpha,cl}(t)} = -i 
\left.\frac{\delta Z_\lambda}{\delta\lambda^q(t)}\right|_{\lambda^q=0}.
\end{equation}
The spin susceptibility is given by a $2$-point function
\begin{equation}
\label{chi}
\chi_{\alpha,\beta}(t,t') = i\av{{\mathcal T_K}S^{\alpha,cl}(t)S^{\beta,q}(t')} = -i 
\left.
\frac{\delta^2 Z_\lambda}{\delta\lambda^{\beta,cl}(t')\delta\lambda^{\alpha,q}(t)}
\right|_{\lambda^q=0},
\end{equation}
while the noise spectrum \cite{kogan} is determined by a different
$2$-point correlator,
$\av{{\mathcal{T}_K}S^{cl}_\alpha(t)S^{cl}_\alpha(t')}$. 

Below in Section ~\ref{sec:BoseKondo}, we will be interested in a
specific $4$-point function that determines the experimentally
accessible noise of susceptibility \cite{Schad14,SendelbachMcDermott2}
\begin{equation}
\label{nos}
C_\chi(t_1,t_1',t_2,t_2') = -
\av{{\mathcal T_K} S^{\alpha,cl}(t_1) S^{\alpha,q}(t'_1) S^{\alpha,cl}(t_2) S^{\alpha,q}(t_2') } = 
-
\left.
\frac{\delta^4Z_\lambda}{\delta\lambda^{\alpha,cl}(t_2')\delta\lambda^{\alpha,q}(t_2) 
                       \delta\lambda^{\alpha,cl}(t_1')\delta\lambda^{\alpha,q}(t_1)} 
\right|_{\lambda^q=0},
\end{equation}
and we focus on the case of identical spin indices.

\subsection{Correlation functions of Majorana fermions in the path-integral representation}
\label{mpath}

Let us now reformulate the mapping between the spin and Majorana
fermion operators discussed above in Sections ~\ref{theorem} and
\ref{auto} in the language of the Keldysh functional integrals. In the
operator language we have established the correspondence (\ref{theo})
between the spin $N$-point functions and Majorana $2N$-point
functions. In special cases (namely, for autocorrelation functions and
`pair' correlators) we found simpler relations (\ref{2CorId}) and
(\ref{4CorId}).

For usual (Dirac) fermions the path integral approach is based on the
concept of coherent states \cite{Kamenev,altland}, which are
eigenstates of the fermionic annihilation operators. An annihilation
operator can only be constructed from {\it two} Majorana fermions. In
many-body problems, this is typically achieved by either ``halving''
\cite{deBoer96,bastianelli,shankar} or ``doubling''
\cite{deBoer96,bastianelli,Nilsson13} the number of Majorana
operators. Keeping in mind applications to systems with a small number
of degrees of freedom (such as the single-spin problem discussed
below), we adopt the doubling procedure, where we add a free Majorana
fermion to each of the operators introduced by the Martin
transformation (\ref{Majrep}). The interaction part of the Keldysh
action can then be formulated in terms of the Grassmann variables
$\eta_\alpha$ corresponding to the Majorana operator $\hat\eta_\alpha$
in the previous sections. The additional Grassmann variables appear
only in the ``free'' quadratic part of the action and are decoupled
from the physical system under consideration.

The Majorana-fermion $2n$-point correlation function on the right-hand
side of Eq.~(\ref{theo}), or rather its time-ordered counterpart, can
be obtained either by $2n$ differentiations of the generating
functional with respect to Grassmann source fields each coupled to a
single Grassmann variable $\eta_\alpha$, or by $n$ differentiations
with respect to $c$-number source fields coupled to {\it pairs} of
Grassmann variables $\eta_\alpha$. These pairs should then be chosen
to represent the spin components according to Eq.~(\ref{Majrep}) as in
(\ref{zs}).  For the autocorrelation functions or the pair correlators
we would like to use the simplified correspondence, i.e.,
Eqs.~(\ref{2CorId}) and (\ref{4CorId}). Here the auxiliary Majorana
fermion $\hat m$ can be represented either by one of the free
Grassmann variables used to construct the functional integral, or by
one out of yet another pair of Grassmann variables that are added to
the theory specifically for the purpose of computing the correlators
(\ref{2CorId}) and (\ref{4CorId}). The source fields can again be
chosen as either Grassmann variables or $c$-numbers. In the explicit
calculation below we chose the latter option. Of course, physical
results are independent of these technical details.

\section{Spin correlators in the Bose-Kondo model}
\label{sec:BoseKondo}

In this section we apply the general conclusion reached in the
previous section to a specific example. For simplicity of the
presentation we use, first, the Matsubara technique, while the final
results are formulated in real time in the frame of the Keldysh
formalism.

As a model we choose a zero-field spin-isotropic Bose-Kondo model (see
Ref.~\cite{ZarandDemlerBoseKondo,ZhuSi} and references therein). This
model appears in various physical contexts, including spin glasses and
liquids~\cite{GeorgesParcolletSachdev2000,GeorgesParcolletSachdev2001}.
In the context of $1/f$ noise a similar model is known as the
Dutta-Horn model \cite{dutta}, where a large number of independent
spins are coupled each to their own bath. In the Majorana
representation introduced above \eqref{Majrep}, the model Hamiltonian
reads
\begin{align}
 H= \vec S \vec X + H_B = -\frac{i}{2} X^\alpha \epsilon_{\alpha\beta\gamma} \eta_\beta \eta_\gamma  + H_B\ ,
\label{Hint}
\end{align}
where $H_B$ is the Hamiltonian of the bosonic bath controlling the
free dynamics of $\vec X$. The latter may be characterized by a
Matsubara correlation function $\langle T
X_\alpha(\tau)X_\beta(\tau')\rangle = \delta_{\alpha \beta}
\Pi(\tau-\tau')$. The function $\Pi(\tau-\tau')$ can be written as
\begin{equation}
\Pi(i\omega_m) = \int\limits_{-\Lambda}^{\Lambda} \frac{dx}{\pi} \,\frac{\rho(|x|)
\sign x}{x - i\omega_m} \ , 
\end{equation}
where $\rho(|x|)$ is the bath spectral density and $\omega_m = 2\pi m
T$. In the Ohmic case considered here, $\rho(|x|) = g |x|$. For
$\omega_m\ll \Lambda$, this gives $\Pi(i\omega_m) \approx
(2g/\pi)\Lambda -g |\omega_m|$. One can perform an RG procedure by
integrating out energies of the bath between $\Lambda/b$ and
$\Lambda$. As a result the coupling constant $g$ is rescaled. The RG
differential equation reads $dg/d\ln b = - 2g^2/\pi$ (see, e.g.,
Ref.~\cite{ZarandDemlerBoseKondo,ZhuSi}). One can supplement this
with the scaling equation for the quasi-particle weight $ d\ln Z/d\ln
b = - 2g/\pi$, which could be important~\cite{PhysRevB.61.15152} as we
are interested in Green's functions of the Majorana fermionic
operators. For $g_0 = g(\ln b=0) \ll 1$, the renormalization effects
are not important as long the the temperature is high $T \gg T_K
\equiv \Lambda \exp[-\pi/(2g_0)]$. Here we assume this to be the
case. Thus we can safely reduce the cutoff $\Lambda(b)$ to a value of
the order of temperature.

\subsection{Matsubara path integral}

We use the Matsubara imaginary-time technique ($t= -i\tau$, $\partial_\tau = -i \partial_t$). 
The partition function ($Z = \int D[\dots] \exp{[iS]}$, for brevity we omit the source fields) reads 
\begin{align}
Z &= \int D[\vec X] D[\eta_\alpha] \, \exp\left\{i \,S_B + i\, \int\limits_0^{1/T} d\tau \left[\frac{1}{2}\,\eta_\alpha(\tau)\,i\, \partial_\tau \eta_\alpha(\tau) + \frac{1}{2}\, X^\alpha(\tau) \epsilon_{\alpha\beta\gamma} \eta_\beta(\tau) \eta_\gamma(\tau) \right] \right\}\ . 
\end{align}
Here $S_B$ is the free bosonic action. 
The first step is to average over the fluctuations of $\vec X$, yielding 
\begin{align}
 Z & = \int D[\eta_\alpha] \, \exp\left\{ -\frac 12 \int d\tau \, \eta_\alpha(\tau) \, \partial_{\tau} \, \eta_\alpha(\tau) - \frac 14 \int d\tau d\tau' M_{\alpha\beta}  \, \Pi(\tau-\tau') \, \eta_{\alpha}(\tau) \eta_{\beta}(\tau) \,  \eta_{\alpha}(\tau') \eta_{\beta}(\tau') \right\}\ ,
\end{align}
the matrix $M$ is of the form
\begin{align}
 M&= \begin{pmatrix} 0 & 1 & 1 \\ 1 & 0 & 1 \\ 1 & 1 & 0 \end{pmatrix}.
\end{align}
Next we decouple the quartic Majorana-interaction in a different channel. To this end we rearrange 
\begin{align}
i S_{M,int}[\eta_\alpha]&= - \frac 14 \int d\tau d\tau' M_{\alpha\beta}  \, \Pi(\tau-\tau') \, \eta_{\alpha}(\tau) \eta_{\beta}(\tau) \,  \eta_{\alpha}(\tau') \eta_{\beta}(\tau')  \notag\\
 &= \frac 14 \int d\tau d\tau'  \, \Pi(\tau- \tau') \, \left[ \eta_{\alpha}(\tau) \eta_{\alpha}(\tau')\right] \, M_{\alpha\beta} 
 \left[\eta_{\beta}(\tau) \eta_{\beta}(\tau')\right] \ .
 \label{Smint}
\end{align}
We now  employ  the Hubbard-Stratonovich transformation by introducing fields $\Sigma_\alpha$. These fields inherit the symmetry 
of Majorana propagators, therefore $\Sigma_\alpha(\tau,\tau')=-\Sigma_\alpha(\tau',\tau)$. The new effective action reads
\begin{align} \label{SetaQMMM}
 i S[\eta_\alpha,\Sigma_\alpha] = & \int d\tau d\tau'  \left(\frac{1}{2}\,\eta_\alpha(\tau) \, (G^{-1}_\alpha)_{\tau\tau'}  \, \eta_\alpha(\tau') 
 - \frac14  \frac{\Sigma_\alpha(\tau,\tau') \,M^{-1}_{\alpha\beta} \,\Sigma_\beta(\tau,\tau')}{\Pi(\tau-\tau')} \right)\ .
\end{align}
The Majorana Green's function in \eqref{SetaQMMM} is
\begin{align}
 (G^{-1}_\alpha)_{\tau\tau'} &= - \delta(\tau-\tau') \partial_{\tau'} - \Sigma_\alpha(\tau,\tau') \ .
 \end{align}

The function $\Pi(\tau-\tau')$ is positive and non-zero. The standard form reads 
\begin{equation}
\Pi(\tau-\tau') = \frac{g \pi T^2}{\sin^2(\pi T (\tau-\tau'))}\ .
\end{equation}
It is cut off at short times, $|\tau-\tau'| < 1/\Lambda$, leading to the maximal value of order $g
\Lambda^2$. 
We use the divergent form keeping the regularization and renormalization in mind. 
Diagonalizing $\hat M^{-1}$, we choose the eigenmodes with positive eigenvalues to be real and
eigenmodes with negative eigenvalues to be imaginary, such that the overall sign of the action
term in the exponent is negative, and the functional integral over $\Sigma_\alpha$ converges.
In other words, we choose $\boldsymbol{\Sigma} = \boldsymbol{\Sigma}' +
i\boldsymbol{\Sigma}''$, where the real part is `diagonal', $\boldsymbol{\Sigma}' =
\Sigma' \cdot(1,1,1)$,
and the imaginary part $\boldsymbol{\Sigma}''$ is orthogonal to it,
(i.e., $\Sigma''_x + \Sigma''_y + \Sigma''_z = 0$ and
$\boldsymbol{\Sigma}''$ describes two degenerate modes); with this choice the 3D integral over
$\Sigma'$ and $\boldsymbol{\Sigma}''$ converges.
%
The redecoupled action \eqref{SetaQMMM} is again quadratic in Majorana Grassmann variables
$\eta_\alpha$,
which allows us to integrate them out and to obtain an effective action of
$\Sigma$-fields:
\begin{align} \label{SQ}
 i S[\Sigma_\alpha]= \frac 12  \sum_{\alpha=\{x,y,z\}} \text{Tr} \log{\left(G^{-1}_\alpha\right)} 
 - \frac14\int d\tau  d\tau'\,   \frac{\Sigma_\alpha(\tau,\tau') \,M^{-1}_{\alpha\beta} \,\Sigma_\beta(\tau,\tau')}{\Pi(\tau-\tau')} \ .
 \end{align}

\subsection{Saddle point solution}

We can now identify the saddle point and fluctuations of the effective $\Sigma$-action.
The saddle-point solution is found by expanding $\Sigma_\alpha=\Sigma_{0\alpha}+\delta
\Sigma_\alpha$. The linear order in 
$\delta \Sigma$ vanishes for
\begin{align}\label{eq:saddlepoint}
\Sigma_{0\alpha}(\tau-\tau')&=\Pi(\tau-\tau') M_{\alpha\beta} G_{0\beta}(\tau-\tau') \,,
\end{align}
where $G_{0\beta}= G_\beta[\Sigma_{0\beta}]$.
A straightforward calculation, in which we disregard the 
broadening of $G_{0\beta}$ on the r.h.s. of Eq.~(\ref{eq:saddlepoint}), leads to 
\begin{equation}
\Sigma_{0\alpha}(i\epsilon_n)  =- \int\limits_{-\Lambda}^{\Lambda}
\frac{dx}{\pi}\,\frac{gx\coth[\beta x/2]}{x-i\epsilon_n}\ .
\end{equation}
Upon the analytic continuation $i\epsilon_n \rightarrow \epsilon + i 0 $ and for $\epsilon
\rightarrow 0$, we obtain the retarded self-energy
\begin{equation}\label{SigmaSaddlePoint}
\Sigma_{0\alpha}^R(\epsilon\rightarrow 0) = - 2i g T = -i\Gamma\ . 
\end{equation}
Here we recognize $\Gamma=2gT$ to be the (Korringa) relaxation rate. In terms of the eigenmodes
this solution means 
$\boldsymbol{\Sigma}''_0=0$, whereas
$\Sigma'_0(i\epsilon_n)=\Sigma_{0\alpha}(i\epsilon_n)$.

\subsection{Fluctuations}
\label{fluct}

To study the fluctuations we expand the \trlog-term in \eqref{SQ} to second order in
$\delta\Sigma$. 
The action reads
\begin{eqnarray}
  \qquad i S_{\delta \Sigma} &=&- \frac{1}{4} \sum_\alpha \int d\tau_1 d\tau_2 d\tau_3 d\tau_4 \, G_{0,\alpha}(\tau_1-\tau_2) \, 
  \delta \Sigma_{\alpha}(\tau_2,\tau_3) 
  G_{0,\alpha}(\tau_3-\tau_4) \, \delta \Sigma_{\alpha}(\tau_4,\tau_1)\nonumber  \\
    &-& \frac14 \int d\tau d\tau' \,\frac{\delta \Sigma_\alpha(\tau,\tau') \,M^{-1}_{\alpha\beta} \,\delta \Sigma_\beta(\tau,\tau')}{\Pi(\tau-\tau')} \ . 
\label{SdSigma}
\end{eqnarray}
The Fourier transform of $\delta\Sigma$ is introduced via
\begin{align}
 \delta\Sigma_\alpha(\epsilon_n,\nu_m)= \int d\tau d\tau ' \, e^{i \nu_m(\tau+\tau')}\, e^{i\epsilon_n(\tau-\tau')} \delta \Sigma_\alpha(\tau,\tau') \ . 
\end{align}
The direct analysis shows that one of $\epsilon_n$ and $\nu_m$ must be fermionic and the
other bosonic, 
so that both $\epsilon_n + \nu_m$ and $\epsilon_n - \nu_m$ are fermionic. Then we obtain 
\begin{eqnarray}
  \qquad i S_{\delta \Sigma} &=&- \frac{T^2}{4} \sum_\alpha  \sum_{\epsilon_n,\nu_m} \, G_{0,\alpha}(\epsilon_n+\nu_m) \, 
   G_{0,\alpha}(\epsilon_n - \nu_m) \, \delta \Sigma_{\alpha}(\epsilon_n,\nu_m) 
\, \delta \Sigma_{\alpha}(\epsilon_n, -\nu_m)\nonumber  \\
    &-& \frac{T^3}{4}\sum_{\alpha\beta} \sum_{\epsilon_1\epsilon_2\nu}\,A(\epsilon_1+\epsilon_2)\,
    \delta \Sigma_\alpha(\epsilon_1,\nu) \,M^{-1}_{\alpha\beta} \,\delta \Sigma_\beta(\epsilon_2,-\nu) \ . 
\label{SdQSigmaFourier}
\end{eqnarray}
Here
\begin{equation}
A(i\omega_m) = \int\limits_0^{1/T} d\tau\, \frac{e^{i\omega_m \tau}}{\Pi(\tau)} = \frac{1}{g\pi T^2} \int\limits_0^{1/T} d\tau\, e^{i\omega_m \tau}\,\sin^2(\pi T \tau) =\frac{1}{2g\pi T^3} \left(\delta_{m,0}-\frac{1}{2}\delta_{m,1}-\frac{1}{2}\delta_{m,-1}\right) \ .
\end{equation}
The mean-field Green function reads
\begin{equation}
G_{0,\alpha}(i\epsilon_n) = \frac{1}{i\epsilon_n +i \Gamma\sign(\epsilon_n)}\ . 
\end{equation} 
Since $\epsilon_n$ is necessarily fermionic, we have $|G_{0,\alpha}|<1/(\pi T)$. Thus, the first term of 
(\ref{SdQSigmaFourier}) cannot compete with the second one which is proportional to $1/g$. This
important observation allows us to disregard the first term of
(\ref{SdQSigmaFourier}) and essentially all the 
contributions of the second and higher orders, originating from the \trlog{}
term of (\ref{SQ}). This in turn simplifies calculations of the higher-order
spin correlators in the next section.

The argument above for the smallness of the first term of 
(\ref{SdQSigmaFourier}) is based on the discreteness of the Matsubara fermionic frequencies. 
Ultimately, we are interested in real times and the behavior in various frequency ranges,
including low frequencies $\omega \ll T$; hence one should be careful with the estimates.
The expressions above indicate that for $\epsilon \rightarrow 0$ the Green functions $G_{0,\alpha}$
in the first term of (\ref{SdQSigmaFourier}) might 
become of order $1/\Gamma = 1/(2gT)$. This, in turn, might imply that the (prefactor of
$\delta\Sigma^2$ in the) first term of (\ref{SdQSigmaFourier}) 
scales with $g^{-2}$ and dominates over the second term $\propto g^{-1}$.
To clarify the situation in the low-frequency range we perform a direct
Keldysh calculation in \ref{sec:pathint_keldysh}. 
We conclude that the first term of (\ref{SdQSigmaFourier}) 
does not become large in the domain of low real 
frequencies. Thus, the second term of (\ref{SdQSigmaFourier}) 
dominates. 

\subsection{Averaging over fluctuations}

The knowledge of the propagator of the $\delta \hat \Sigma$-fluctuations allows one to construct a
new perturbative series, starting at the fixed-point solution given in Eq.~\eqref{SigmaSaddlePoint}.
The new perturbative expansion for the Green function is based on the following series:
\begin{multline}
 \hat G_{\alpha}(\tau,\tau') = \hat G_{0,\alpha}(\tau,\tau') + \int d\tau_1 d\tau_2 \hat G_{0,\alpha}(\tau,\tau_1) \delta\hat{\Sigma}_\alpha(\tau_1,\tau_2) \hat G_{0,\alpha}(\tau_2,\tau') \\ 
  + \int d\tau_1 d\tau_2 d\tau_3 d\tau_4 \hat G_{0,\alpha}(\tau,\tau_1) \delta\hat{\Sigma}_\alpha(\tau_1,\tau_2) \hat G_{0,\alpha}(\tau_2,\tau_3) \delta\hat{\Sigma}_\alpha(\tau_3,\tau_4) \hat G_{0,\alpha}(\tau_4,\tau') +...
\label{Gexp}
\end{multline}
This series is to be averaged over the fluctuations to obtain $\left\langle\hat
G_\alpha(t,t')\right\rangle_{\delta \Sigma}$. However, on the right-hand side the
spin index $\alpha$ is the same in all the terms, essentially because the
saddle-point Green function is diagonal in spin space. Taking into account the dominance of the
second term in the action (\ref{SdQSigmaFourier}), we conclude that 
\begin{align}
 \av{\delta\hat\Sigma_\alpha(\dots)\delta\hat\Sigma_\alpha(\dots)}\propto M_{\alpha\alpha}=0\ ,
\label{Sigcor}
\end{align}
because the diagonal entries of the matrix $\hat M$ are zeros (note that in Eq.~\eqref{Sigcor}
the contributions of the real and imaginary $\Sigma$-components cancel each other,
cf.~the discussion above Eq.~\eqref{SQ}). Thus, the fluctuation-averaged Green function coincides
with the saddle-point solution
\begin{align}
 \left< \hat G_\alpha(t,t')\right>_{\delta \Sigma}= \hat G_{0,\alpha}(t,t') \ .
\label{Gphys}
\end{align}
Moreover, for averages of Green-function products with the same spin index
$\alpha$ we can substitute all $\hat G_\alpha$ with their saddle-point values due to
the relation (\ref{Sigcor}); this simplifies calculations considerably. In the next section
we employ this convenient property. 

\subsection{Correlation functions}
\label{corr}

\begin{figure}
\centering
\includegraphics[width=.2\textwidth]{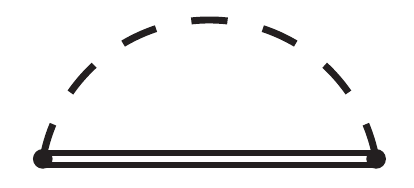}
\caption{The diagram for spin susceptibility.} 
\label{fig:Susceptibility}
\end{figure}

As a first example, we calculate the diagonal spin susceptibility
\begin{eqnarray}
\chi(t,t') = i\av{{\mathcal T_K}S^{\alpha,cl}(t)S^{\alpha,q}(t')}=
-i \av{{\mathcal T_K}\left[\hat{m}(t)\hat\eta_\alpha(t)\right]^{cl}\left[\hat{m}(t')\hat\eta_\alpha(t')\right]^{q} } \ .
\end{eqnarray}
The only contribution to $\chi(t,t')$ is given by the diagram in Fig.~\ref{fig:Susceptibility}.
We immediately arrive at the standard result 
\begin{equation}
\chi''(\omega)=\frac{1}{2}\frac{\Gamma\tanh\frac{\omega}{2T}}{\omega^2+\Gamma^2} 
\overset{\omega\ll T}\approx \frac{1}{4T}\frac{\omega \Gamma}{{\omega^2+\Gamma^2} }\ .
\end{equation}
Notice that no vertex corrections appear in our case, and the precision of this calculation rests
solely on the precision, with which the self-energy is evaluated. 

\begin{figure}
\centering
\showgraph{width=.2\textwidth}{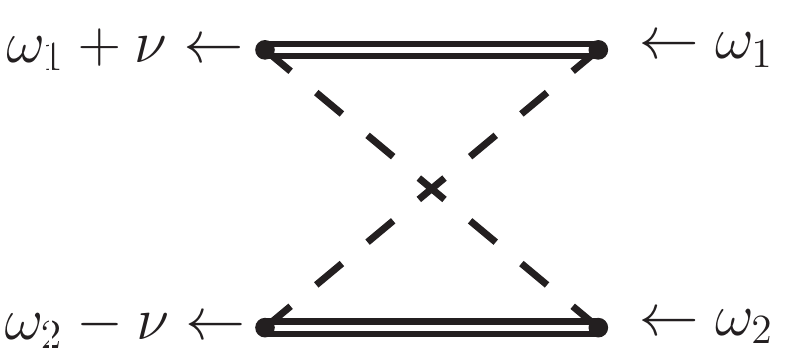} + \showgraph{width=.1\textwidth}{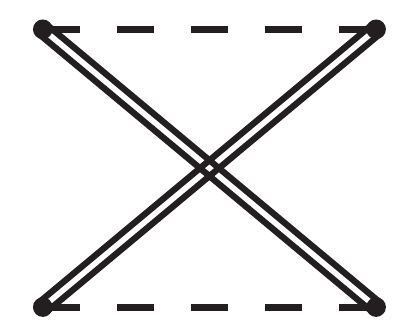} 
+\showgraph{width=.1\textwidth}{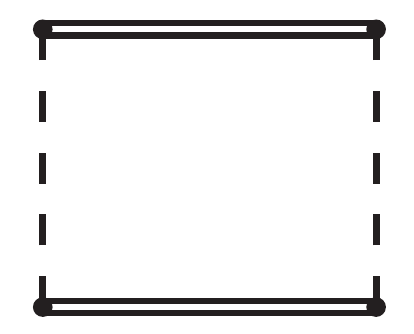} + \showgraph{width=.1\textwidth}{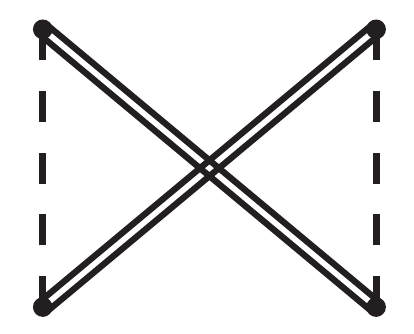} 
+ \showgraph{width=.1\textwidth}{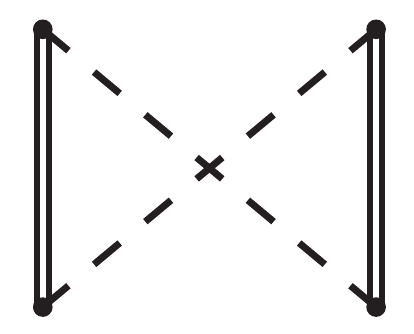} + \showgraph{width=.1\textwidth}{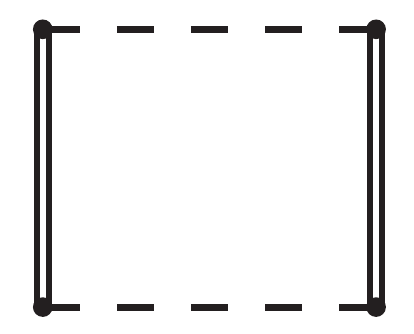} 
\caption{The 'lowest order' diagrams to non-equilibrium noise of susceptibility.} 
\label{fig:Cchi0}
\end{figure}

We further demonstrate the power of our approach by calculating one of the higher-order spin
correlators. We evaluate the
connected part of the 4-th order correlator (\ref{nos}), which is related to noise of spin susceptibility~\cite{Schad14,SendelbachMcDermott2}.
Using (\ref{4CorId}) we obtain
\begin{eqnarray}
C_\chi(t_1,t_1',t_2,t_2') &=& -
\av{{\mathcal T_K} S^{\alpha,cl}(t_1) S^{\alpha,q}(t'_1) S^{\alpha,cl}(t_2) S^{\alpha,q}(t_2') } \nonumber\\&=&-
\av{{\mathcal T_K}\left[\hat{m}(t_1)\hat\eta_\alpha(t_1)\right]^{cl}\left[\hat{m}(t_1')\hat\eta_\alpha(t_1')\right]^{q} 
            \left[\hat{m}(t_2)\hat\eta_\alpha(t_2)\right]^{cl}\left[\hat{m}(t_2')\hat\eta_\alpha(t_2')\right]^{q}}\ .
\end{eqnarray}
The discussion in the previous section implies that the connected part of this
correlator is given by the six diagrams 
depicted in Fig.~\ref{fig:Cchi0}. Here, the double solid lines stand for the saddle-point Green
functions $\hat G_{0,\alpha}$, whereas 
the dashed lines represent the trivial correlators of the conserved quantity $\hat{m}$. As shown in
the previous section, we may use the saddle-point Green functions, since the contribution of the
$\delta \hat\Sigma_\alpha$-fluctuations vanishes, cf.~(\ref{Sigcor}).
For completeness, we show the result, which is put into context and interpreted in detail
elsewhere~\cite{Schad14}.
\begin{eqnarray} 
\label{res1}
&& 
C^{conn}_\chi(\nu,\omega_1,\omega_2) = \frac{i\Gamma^2}{8T^2} \frac{\omega_1+\omega_2 + 2 i \Gamma}
{(\omega_1+i\Gamma)(\omega_2+i\Gamma)(\omega_1+\nu+i\Gamma)(\omega_2-\nu+i\Gamma)}.
\nonumber
\end{eqnarray}

If a spin correlation function involves different spin components, one can no longer rely on the 
saddle-point contributions. One example is the correlator
\begin{align}
  \av{{\mathcal T_K} S^{\alpha,cl}(t_1) S^{\alpha,q}(t'_1) S^{\beta,cl}(t_2) S^{\beta,q}(t_2') } \,,
\end{align}
related to the correlations of susceptibilities in different directions.
The non-zero off-diagonal fluctuations $\av{\delta\hat
\Sigma_\alpha \delta\Sigma_\beta}$ contribute to this
correlator, and thus additional diagrams have to be
considered. This is, however, beyond the scope of the present paper.

\section{Conclusion} 
\label{sec:conc}

In this paper we have demonstrated the efficiency of the Majorana
representation of the spin-$1/2$ operators (\ref{Majrep}). We have 
shown that the representation (\ref{Majrep}) allows for a direct
calculation of spin correlation functions in terms of Majorana
fermions despite the fact that their Hilbert space is enlarged as
compared to that of the spins. The precise construction of the
Majorana Hilbert space was shown to be irrelevant as far as the
correlation functions are concerned. 

The Majorana representation is particularly efficient in the case of
auto-correlation functions (or ``pair-wise'' correlation functions).
Such $N$-spin functions can be represented in terms of $N$-point
Majorana correlators, which significantly simplifies calculations. 
In particular complicated vertex structures do not appear.  

As an example we have revisited the well known Bose-Kondo model. 
We have developed the Keldysh path-integral approach and have shown that 
the spin relaxation and susceptibility are efficiently described within the  
saddle-point approximation. Moreover we have shown that correlation functions 
containing a single spin projection can also be efficiently calculated at the saddle-point. 
In particular we have evaluated a $4$-spin correlation function corresponding to the 
noise of susceptibility. 

It would be interesting to apply our approach to a wider range of physical problems, for example, 
to a sub-Ohmic Bose-Kondo model describing physics of spin glasses~\cite{GeorgesParcolletSachdev2000,GeorgesParcolletSachdev2001,ZarandDemlerBoseKondo,ZhuSi}.

\section{Acknowledgements} 

We acknowledge discussions with A. Mirlin and J. Clarke. 
This research was funded by the Intelligence Advanced Research Projects 
Activity (IARPA) through the US Army Research Office, the German 
Science Foundation (DFG) through Grants No. SCHO 287/7-1 and No. SH 81/2-1, 
the German-Israeli Foundation (GIF), and EU  Network  Grant  Inter-
NoM.

\appendix

\section{Perturbation theory}
\label{pert}

\begin{figure}
\centering
\includegraphics[scale=1.1]{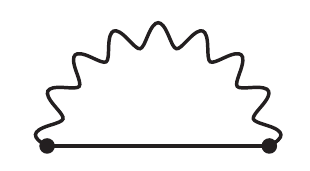}
\caption{The leading contribution to the self-energy.}
\label{fig:SelfEnergy}
\end{figure}

Given the Majorana representation for spins \eqref{Majrep}, the
identities \eqref{2CorId},
\eqref{4CorId}, and the Hamiltonian \eqref{Hint}, one could calculate
correlation functions within
perturbation theory by diagrammatic expansion. Here the zeroth-order
Hamiltonian
consists of the bath part $H_B$ only. The dressed Majorana Green
function is self-consistently
constructed, using the lowest-order self-energy. The lowest-order
self-energy
diagram is depicted in Fig. \ref{fig:SelfEnergy}, it contains the
``free'' Majorana Green
function $\hat G_{f,\alpha}(\omega)$,
\begin{gather}
G_{f,\alpha}^{R/A}(\omega)= \left( \omega \pm i 0 \right)^{-1},\\
\Sigma^R_{\alpha}(\omega)= - \int\frac{d\Omega}{2\pi} \left(
\Pi^K(\omega+\Omega)
G_{f,\alpha}^A(\Omega) + \Pi^R(\omega+\Omega) G_{f,\alpha}^K(\Omega)
\right) = -2 i g T \left(1  + \mathcal{O} \left( \frac{\omega}{T}
\right) \right), \label{PtSigma}\\
G_\alpha^R(\omega) = \left( \omega - \Sigma^R_{\alpha}(\omega)
\right)^{-1} = \left( \omega +i\Gamma \right)^{-1},\qquad \Gamma=
2 g T.
\label{PtG}
\end{gather}
Here the Keldysh components of the bath correlation functions read
\begin{gather}
\hat\Pi^{ab}_{\alpha\beta}(t,t')= \delta_{\alpha\beta}
\av{\mathcal{T}_K \hat X^{\alpha,a}(t) \hat X^{\alpha,b}(t')} =
\hat\Pi^{ab}(t,t'), \\
 \Pi^{R/A}(\omega) = \pm g \omega -i D,\qquad \Pi^K(\omega)=
\coth\left(\frac{\omega}{2 T}\right) \left( \Pi^R(\omega) -\Pi^A(\omega)
\right) \ ,
\end{gather}
where $D\equiv (2g/\pi)\Lambda$, the spin indices
$\alpha,\beta=\{x,y,z\}$, the Keldysh indices $a,b=\{cl,q\}$, and the
"classical" and "quantum" operators $\hat X^{\alpha,cl(q)}=(\hat
X^{\alpha,u} \pm X^{\alpha,d})/\sqrt{2}$.

\begin{figure}
\centering
\showgraph{width=.09\textwidth}{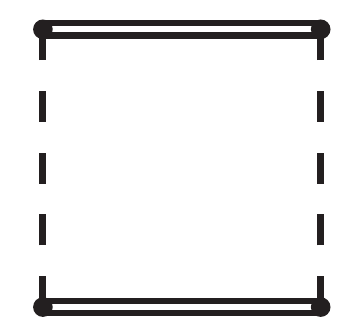} +
\showgraph{width=.15\textwidth}{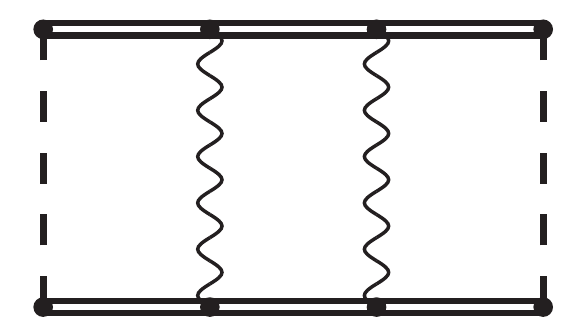} +
\showgraph{width=.15\textwidth}{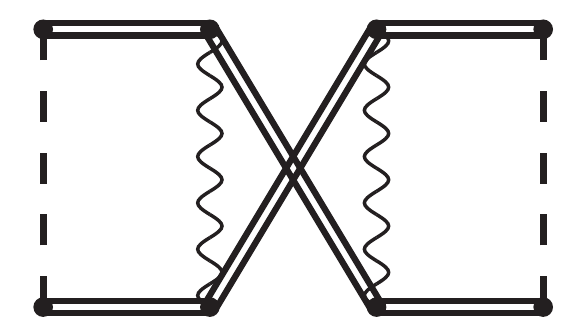} +...
\caption{Examples of diagrams, which contain connected Majorana propagators.}
\label{fig:C1cor}
\end{figure}

\begin{figure}
\showgraph{width=.3\textwidth}{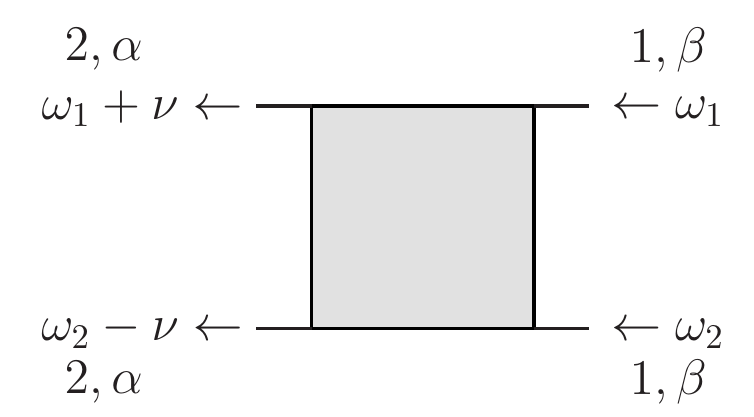} = \showgraph{width=.065\textwidth}{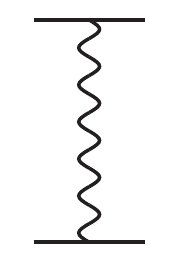} 
+ \showgraph{width=.22\textwidth}{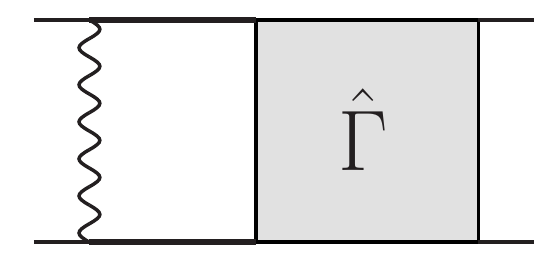} 
+ \showgraph{width=.24\textwidth}{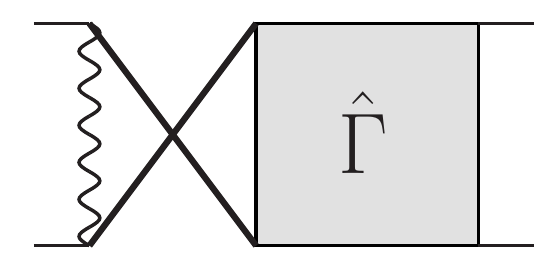} 
\caption{Dyson equation for the partially dressed bosonic interaction. Thick solid lines are dressed
Majorana propagators, wavy lines correspond to the bosonic bath.} 
\label{fig:RenInt}
\end{figure}

Typically, this method is used to calculate 1- or 2-point correlation functions, e.g.,
magnetization or susceptibility. In lower orders Majorana propagators are never connected by a
bosonic line. Such an element only appears in higher orders or in higher correlation functions as
discussed in \cite{Schad14}. An example is shown in Fig.~\ref{fig:C1cor}. Here it is in principle
necessary to consider more complex diagrams including, e.g., ladders of bosonic propagators. 
In the perturbation theory with a vanishing unperturbed spin Hamiltonian it is by no 
means justified to simply neglect these diagrams.
As an illustration let us consider renormalization of the interaction line. The dressed interaction
carries four times (or frequencies), the Keldysh indices
$a,b,c,d=\{cl,q\}$, and the spin indices $\alpha,\beta,\gamma,\delta=\{x,y,z\}$; diagrammatically it
is depicted by
\begin{align}
 \hat\Gamma_{\alpha\beta,\gamma\delta}^{ab,cd} = \showgraph{width=.2\textwidth}{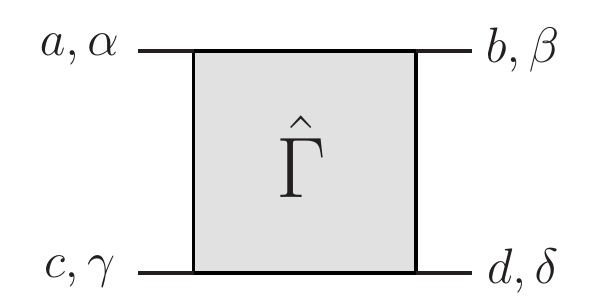}.
\end{align}
For demonstration, we pick one of several possible components of $\hat\Gamma$ that carries
identical spin indices on the left (as well as on the right) and the Keldysh indices (21,21), which
correspond to the retarded-retarded component, that is $\Gamma^{RR}_{\alpha,\beta}$. The
partially dressed interaction is obtained from the Dyson-type equation, depicted in 
Fig.~\ref{fig:RenInt}, by summation of the leading contributions in the small-$g$ expansion. These
are constructed out of the bosonic correlator $\Pi^K\sim 2g T=\Gamma$ combined with Green's functions $G^R G^A$ such that the upper and lower halves of the complex plane each contain one of the Green-functions poles.
These leading contributions are of the same order in $g$ as the bare bosonic line, thus we might suspect 
a strong renormalization of the  interaction line. This is, however, not the case due to a cancellation. 
Taking into account both contributions depicted in Fig.~\ref{fig:RenInt} we obtain 

\begin{multline}
 \Gamma_{\alpha\beta}^{RR}(\omega_1,\omega_2,\nu) = \Pi^K(\nu) M_{\alpha\beta} -  \int \frac{d\Omega}{2\pi}\,\Pi^K(\nu-\Omega) M_{\alpha\gamma} G_\gamma^R(\omega_1+\Omega) G_\gamma^R(\omega_2-\Omega)  \Gamma^{RR}_{\gamma\beta}(\omega_1,\omega_2,\Omega) \\
  + \int \frac{d\Omega}{2\pi}\, \Pi^K(\nu-\Omega) M_{\alpha\gamma} G_\gamma^R(\omega_1+\Omega) G_\gamma^R(\omega_2-\Omega)  \Gamma^{RR}_{\gamma\beta}(\omega_1,\omega_2,-\Omega-\omega_1+\omega_2)
\label{RenInt}
\end{multline}
The Green functions are calculated within the approximation $\Sigma^R(\omega)=-2 i g T=-i\Gamma$, which is justified and consistent in the
high-temperature regime $T\gg T_K,\nu,\omega_1,\omega_2,\Gamma$.
Let us assume that the leading term
in $\Gamma^{RR}$ does not depend on the third frequency and splits $\Gamma^{RR}=
\Gamma_0^{RR} + \Gamma^{RR}_\nu$. If this is the case, the integrals in the second and third term
become equal, and we find that the renormalized interaction coincides with the bare one:
\begin{align}
 \Gamma_{\alpha\beta}^{RR}(\omega_1,\omega_2) = \Pi^K(0) M_{\alpha\beta}.
\end{align}

The problem in this perturbative approach is that the non-perturbed spin Hamiltonian is zero 
(the non-perturbed Hamiltonian consists of the bath Hamiltonian $H_B$ only). In
other words, 
the spin-bath interaction is not weak as compared to the energy scale of the unperturbed spin dynamics. 
One might therefore expect that higher-order diagrams are important. 
Having observed a particular cancellation we cannot in principle exclude other important higher-order 
contributions. Instead of evaluating multiple higher-order
diagrams we choose the path-integral approach in the following section allowing for a more
straightforward analysis.

\section{Majorana path integral in the Keldysh representation} 
\label{sec:pathint_keldysh}

In Section~\ref{sec:BoseKondo} we have developed a path-integral technique, which allowed us to
obtain 
the self-energy (Korringa relaxation rate) as a saddle-point solution. We have argued that the
fluctuations 
around this saddle point can sometimes be disregarded. Namely, this is the case for a correlation
function involving only one spin component.
This conclusion was based on the smallness of the higher-than-linear contributions to the
\trlog{} term of the action~(\ref{SQ}). We have shown this using the small parameter
$\Gamma/\epsilon_n$ for the fermionic Matsubara frequencies. To be able to treat also low real
frequencies, we provide here the Keldysh version of the path-integral calculation.
  
For the model considered here, \eqref{Hint}, the partition function reads
\begin{align}
 Z &= \int D[\vec X]D[\eta_\alpha] \, \exp\left\{iS_B+ \frac i2 \int_C dt \left( \eta_\alpha(t) i\partial_t \eta_\alpha(t)+ i X^\alpha(t) \epsilon_{\alpha\beta\gamma} \eta_\beta(t) \eta_\gamma(t) \right) \right\} \nonumber \\
  &= \int D[\vec X] D[\eta_\alpha] \, \exp\left\{iS_B + \sum_{a=u,d} \frac i2 \int^\infty_{-\infty} dt \, \left( \eta^a(t) i\tau_3^a \partial_t \eta^a(t)+ i \tau_3^a X^{a,\alpha}(t) \epsilon_{\alpha\beta\gamma} \eta^a_\beta(t) \eta^a_\gamma(t) \right) \right\}\ ,
\end{align}
where the Keldysh index $a$ takes the value $u$ at the forward part of the contour and
$d$ on the backward part of the contour. 
At the first step we average over the fluctuations of $\vec X$, which yields
\begin{align}
 Z & = \int D[\eta_\alpha] \, \exp\left\{ -\frac 12 \int dt \, \eta^a_\alpha(t) \, \tau_3^{ab} \partial_{t} \, \eta^{b}_\alpha(t) + \frac 14 \int dtdt' M_{\alpha\beta} \tau_3^a \, \Pi^{ab}(t,t') \,\tau_3^b \  \eta^a_{t,\alpha} \eta^a_{t,\beta} \,  \eta^b_{t',\alpha} \eta^b_{t',\beta} \right\}\ .
\end{align}
Here $\langle T X^a_\alpha(t)X^b_\beta(t')\rangle = \delta_{\alpha \beta} \Pi^{ab}(t-t')$, and $a,b$
are the Keldysh indices over which summation is implied. As in Section~\ref{sec:BoseKondo}, we
decouple the quartic term in a different channel
\begin{align}
  i S_{M,int}[\eta_\alpha]&= \frac 14 \int dtdt' M_{\alpha\beta} \tau_3^a \, \Pi^{ab}(t,t') \,\tau_3^b \  \eta^a_{t,\alpha} \eta^a_{t,\beta} \  \eta^b_{t',\alpha} \eta^b_{t',\beta}  \notag\\
 &= -\frac 14 \int dtdt' \tau_3^a \, \Pi^{ab}(t,t') \,\tau_3^b \ ( \eta^a_{t,\alpha} \eta^b_{t',\alpha}) \, M_{\alpha\beta} ( \eta^a_{t,\beta} \eta^b_{t',\beta}).
\label{Smint_app}
\end{align}

\subsection{Qualitative considerations}

Prior to performing the full fledged Keldysh analysis, we provide here a qualitative argument based
on the locality 
of the bath correlation function $\Pi^{ab}(t-t')$ on the relevant time scale of order $1/\Gamma$. On this time-scale we can 
safely replace all components of $\Pi^{ab}(t-t')$ by its classical (Keldysh) part, i.e., $\Pi^{ab}(t-t') \approx \tilde \Pi(t_1-t_2)=(1/2)\Pi^K(t-t')$. 
This allows us to proceed similarly to the treatment in the Matsubara case in the main text
(cf.~Eq.~\ref{SQ}), and we obtain 
\begin{align} \label{SQKapprox}
 i S[\Sigma_\alpha]= \frac 12  \sum_{\alpha=\{x,y,z\}} \text{Tr} \log{\left(G^{-1}_\alpha\right)} 
 - \frac14\int d t  d t'\,   \frac{\Sigma^{ab}_\alpha(t,t') \,\tau_3^{a} \tau_3^{b}
M^{-1}_{\alpha\beta} \,\Sigma^{ab}_\beta(t,t')}{\tilde\Pi(t-t')} \,,
 \end{align}
 where 
 $$
 (G_\alpha^{-1})^{ab}_{tt'} = i \tau_3^{ab} \delta(t-t') \partial_{t'} - \Sigma^{ab}_\alpha(t,t')\ .
 $$
 One can find the saddle point and again obtain the relaxation rate $\Gamma = 2gT$. This
is done below 
 in the full Keldysh calculation. Here we concentrate on the fluctuations $\delta \Sigma^{ab}$.
 On the relevant time scales ($\sim \Gamma^{-1}$) the function $\tilde \Pi$ is local, $\tilde
\Pi(t-t') \sim 2g T \delta(t-t')$. 
 (Note that the delta-function should be understood as such only at relatively long time scales.
For instance, it does not force us to take the Grassmann variables in~(\ref{Smint_app}) at
coinciding times.) This locality means, in turn, that 
 the fluctuating self-energies $\delta \Sigma^{ab}$ are local. On the other hand they are anti-symmetric, $\Sigma^{ab}(t,t') 
 = - \Sigma^{ba}(t',t)$. This strong constraint implies that only off-diagonal (in Keldysh indices)
components of $\delta\Sigma$ fluctuate: $\delta\Sigma^{ud}(t,t')= - \delta\Sigma^{du}(t',t)
\sim \delta(t-t')$, whereas $\delta\Sigma^{uu}=\delta\Sigma^{dd}=0$. Upon the Keldysh rotation (see
below), this means that only the retarded and advanced components of 
 $\delta \hat \Sigma$ fluctuate. In addition the Keldysh component of the Majorana Green function 
 $G_\alpha^K$ can be neglected, since it scales as $\propto \tanh(\omega/2T) \delta(\omega)$. As a
result, upon expansion of the \trlog{} term of~(\ref{SQKapprox}), the only terms that can
appear are of the type 
 $\text{Tr}[G_\alpha^R \delta\Sigma^R G_\alpha^R \delta\Sigma^R G_\alpha^R \delta\Sigma^R \dots]$ and
 $\text{Tr}[G_\alpha^A \delta\Sigma^A G_\alpha^A \delta\Sigma^A G_\alpha^A \delta\Sigma^A \dots]$. 
 Since $\delta \Sigma^{R/A}$ are local in time, these terms vanish. Thus we are allowed to disregard
the \trlog{} term of~(\ref{SQKapprox}). Notice, that this argument is not valid at finite
magnetic fields ($B \geq \Gamma$, see below).

\subsection{Full Keldysh calculation}

We now go back to the full Keldysh version of~(\ref{SQKapprox}) keeping all Keldysh components of
$\Pi^{ab}$.
One can decouple the quartic Majorana interaction with the help of complex bosonic fields $Q_\alpha$
via a Hubbard-Stratonovich transformation. The fields $Q_\alpha$ inherit the symmetry of the
Majorana propagators, therefore $Q^{ab}_\alpha(t,t')=-Q_\alpha^{ba}(t',t)$. Since the correlator
$\Pi^{ab}(t-t')$ may have a complicated time-dependent structure, we choose to keep it in the
numerator:
\begin{align} \label{SetaQ}
 i S[\eta_\alpha,Q_\alpha] = & \int dtdt'  \left(\frac{i}{2}\eta^a_\alpha(t) \, (G_\alpha^{-1})^{ab}_{tt'}  \, \eta^{b}_\alpha(t') 
 - \frac14 \tau_3^a \tau_3^b \Pi^{ab}(t,t') \, Q^{ab}_\alpha(t,t') \,M^{-1}_{\alpha\beta} Q^{ab}_\beta(t,t')\right) 
\end{align}
The Majorana Green function in \eqref{SetaQ} reads
\begin{align}
 (G_\alpha^{-1})^{ab}_{tt'} &= i \tau_3^{ab} \delta(t-t') \partial_{t'} - \Sigma^{ab}_\alpha(t,t')\ , \\ 
 \Sigma^{ab}_\alpha(t,t') &=   \tau_3^a \tau_3^b Q_\alpha^{ab}(t,t') \Pi^{ab}(t,t')\ .
\end{align}

After the standard Keldysh rotation,
\begin{gather}
 \hat G=L G L = \begin{pmatrix} G^K & G^R \\ G^A &0 \end{pmatrix},\quad \hat \Pi=L \Pi L=\begin{pmatrix} \Pi^K & \Pi^R \\ \Pi^A &0 \end{pmatrix},\\
 \hat\Sigma=L \Sigma  L= \begin{pmatrix} 0 & \Sigma^A \\ \Sigma^R & \Sigma^K \end{pmatrix},\quad
L=\frac{1}{\sqrt{2}} \begin{pmatrix}1&1\\1&-1\end{pmatrix}\,,
\end{gather} 
we find
\begin{align}
 (\hat G_\alpha^{-1})_{tt'} &= i \tau_1 \delta(t-t') \partial_{t'} - \hat\Sigma_\alpha(t,t')\ , \\
 \hat\Sigma^{ab}_\alpha(t,t') &=  -\frac{1}{2}   \text{Tr}\left\{ \bar\gamma^a \hat  \Pi(t,t')\bar\gamma^b \hat Q_{\alpha}(t',t) \right\}\ ,
\end{align}
where $\bar\gamma^{cl}\equiv\tau_1$ and $\bar\gamma^{q}\equiv\tau_0$.

The re-decoupled action \eqref{SetaQ} is again quadratic in Majorana Grassmann variables
$\eta_\alpha$, enabling us to integrate them out and to obtain an effective action for the
$Q$-fields. Thereafter, we can identify the saddle point and fluctuations of the effective
$Q$-action.
\begin{align} \label{SQKeldysh}
 i S[Q_\alpha]= \frac 12 \text{Tr}_t \sum_\alpha^{x,y,z} \log{(\hat G_\alpha(\hat Q_\alpha))^{-1}} + \frac{1}{8}\int dt dt' \hat \Pi^{ab}(t,t') M^{-1}_{\alpha\beta}\text{Tr}\left\{ \bar\gamma^a \hat Q_\alpha(t,t')\bar\gamma^b \hat Q_\beta(t',t) \right\}.
\end{align}
Here $\text{Tr}_t$ denotes the trace in the Keldysh and time space and $\text{Tr}$ is the trace in
the Keldysh space. The saddle-point solution is found by expansion taking the linear order in
$\delta \hat Q$. The solution  must be stationary, depending only on the time difference: $\hat
G_{0,\beta}(t,t')=\hat G_{0,\beta}(t-t')$. We obtain the self-consistency equation 
\begin{align}
 \hat Q_{0,\alpha}(t-t')&= M_{\alpha\beta}\hat G_{0,\beta}(t-t')\ .
 \label{SPsol} 
\end{align}
In the high-temperature regime, $T\gg T_K$, it is easy to obtain the self-consistent solution for
the self-energy. In frequency space, one finds (summation over double indices assumed)
\begin{align}
 \Sigma^R_{0,\alpha}(\omega)&= -\frac{1}{2} M_{\alpha\beta} \int\frac{d\Omega}{2\pi} \left( \Pi^K(\omega+\Omega) 
 G^A_{0,\beta}(\Omega) + \Pi^R(\omega+\Omega) G^K_{0,\beta}(\Omega) \right)  \notag\\
 &= -2 i g T \left(1 + \mathcal{O} \left( \frac{\omega}{T} \right) \right) 
\label{SPSigma} \\
G_{0,\alpha}^R(\omega) &= \left(\omega -\Sigma_{0,\alpha}^R(\omega)\right)^{-1} \approx \left(\omega + 2i g T \right)^{-1}   \ ,
\label{SPG}
\end{align}
coinciding with the saddle-point result \eqref{SigmaSaddlePoint} and the results of the perturbation
theory \eqref{PtSigma} and \eqref{PtG}.

To analyze the fluctuations $\delta\hat Q$, we expand the \trlog-term in \eqref{SQKeldysh} up to
the
second order in $\delta\hat Q$. With the help of the Fourier transform of $\delta \hat Q$,
introduced as
\begin{align}
 \delta\hat Q_\alpha(\omega,\nu)= \int dt dt' \, e^{-i\frac{\nu}{2}(t+t')}\, e^{-i\omega(t-t')}
\delta\hat Q_\alpha(t,t')\,,
\end{align}
we rewrite the action in the form $ i S_{\delta Q} = iS_{\delta Q}^{(1)} +  iS_{\delta Q}^{(2)}$.
The first term originates in the expansion of the \trlog{} term of (\ref{SQKeldysh}):
\begin{multline}
  \qquad i S_{\delta Q}^{(1)} = \frac{1}{2} \int dt dt'dt_1 dt_1' \, \text{Tr}\left\{ \hat G_{0,\alpha}(t,t_1') \, \delta\hat \Sigma_{\alpha}(t_1',t_1)  \hat G_{0,\alpha}(t_1,t') \, \delta\hat \Sigma_{\alpha}(t',t) \right\} \\
  = \frac{1}{2} \int \frac{ d\nu  d\Omega}{(2\pi)^2} \,\text{Tr}\left\{ \hat G_{0,\alpha}(\Omega-\nu/2) \, \delta\hat \Sigma_{\alpha}(\Omega,\nu)  \hat G_{0,\alpha}(\Omega+\nu/2) \, \delta\hat \Sigma_{\alpha}(\Omega,-\nu)  \right\}\ .
 \label{SdQ1}
 \end{multline}
Here, for brevity, we expressed $i S_{\delta Q}^{(1)}$ via the self-energy fluctuations
\begin{align}
 \delta\hat \Sigma_{\alpha}^{ab}(\Omega,\nu)&= -\frac 12 \int\frac{d\omega}{2\pi} \text{Tr}\left\{ \bar\gamma^a \hat  \Pi(\Omega+\omega) \bar\gamma^b \delta \hat Q_{\alpha}(\omega,-\nu) \right\}  \ .
 \label{dSigma}
\end{align}
The second term appeared after the Hubbard-Stratonovich transformation:
\begin{multline}
  \qquad i S_{\delta Q}^{(2)} = \frac{M^{-1}_{\alpha\beta}}{8}\int dt dt' \hat \Pi^{ab}(t,t') \text{Tr}\left\{ \bar\gamma^a \delta\hat Q_\alpha(t,t')\bar\gamma^b \delta\hat Q_\beta(t',t) \right\} \\
 =  \frac{M^{-1}_{\alpha\beta}}{8}\int \frac{d\nu d\omega d\omega'}{(2\pi)^3} \hat \Pi^{ab}(\omega'-\omega) \text{Tr}\left\{ \bar\gamma^a \delta\hat Q_\alpha(\omega,\nu)\bar\gamma^b \delta\hat Q_\beta(\omega',-\nu) \right\}  \ . 
\label{SdQ2}
\end{multline}

In this Appendix we focus on the first term \eqref{SdQ1}, which emerged from the
expansion of the \trlog{} term. We give a detailed discussion of this term and conclude that it
may be neglected as
compared to the second term \eqref{SdQ2}. 
This is done on the basis of a small-$g$ expansion, justifying our course of action in
Section~\ref{fluct} of the main text.
In order to symmetrize and simplify the problem we parametrize fluctuations around the saddle point in terms 
of the mode $R^{(1)}$, symmetric with respect to time, and three antisymmetric modes $R^{(2)},
R^{(3)}$ and $R^{(4)}$:
\begin{gather}
 \delta\hat Q_{\alpha}(\omega,\nu) =  \begin{pmatrix} R^{(3)}_\alpha(\omega,\nu) + R^{(4)}_\alpha(\omega,\nu) & i R^{(1)}_\alpha(\omega,\nu) + R^{(2)}_\alpha(\omega,\nu) \\ -i R^{(1)}_\alpha(\omega,\nu) + R^{(2)}_\alpha(\omega,\nu) & R^{(3)}_\alpha(\omega,\nu) - R^{(4)}_\alpha(\omega,\nu) \end{pmatrix}  
 \label{dQdecomp}\\
 R^{(1)}_\alpha(t,t')= R^{(1)}_\alpha(t',t),\qquad R^{(2)}_\alpha(t,t')=-R^{(2)}_\alpha(t',t),\qquad R^{(3)}_\alpha(t,t')=-R^{(3)}_\alpha(t',t), \qquad R^{(4)}_\alpha(t,t')=-R^{(4)}_\alpha(t',t) 
\label{dQsym}
\end{gather}

Clearly, (\ref{SdQ2}) is proportional to $g$ essentially originating from the bath correlation
function $\hat\Pi$.
In (\ref{SdQ1}), each $\delta \hat \Sigma$ contains a factor of $\hat\Pi$, therefore the whole term appears to be of at least second order in $g$ unless the Green's functions yield an inverse factor $g$. The only combination of Green's functions yielding $1/g$ is $G^A(\Omega+\nu/2) G^R(\Omega-\nu/2)$ (or vice versa, $R/A\rightarrow A/R$). For example, one of the contributions to~\eqref{SdQ1} has the form
\begin{align}
 \int \frac{d\omega d\omega' d\nu d\Omega}{(2\pi)^4} \, G^A(\Omega+\nu/2) G^R(\Omega-\nu/2) \Pi^K(\Omega+\omega) \Pi^K(\Omega+\omega') R^{(i)}_\alpha(\omega,\nu) R^{(j)}_\alpha(\omega',-\nu), \qquad i,j=3,4  \ .
 \label{DivM}
\end{align}
In this term, assuming $\Pi^K\approx 4g T$ to be constant (at low frequencies), the $\Omega$
integration of $G^R G^A$ yields an inverse factor of $g$ since $1/\Gamma=(2g T)^{-1}$. The structure
of this term resembles the structure of the diagrammatic elements \eqref{RenInt}
discussed~\ref{pert}. However, in writing \eqref{DivM} we did not take into account the antisymmetry
of $R^{(3)}$ and $R^{(4)}$ as explained in \eqref{dQsym}. Due to the antisymmetry terms of the kind
\eqref{DivM} cancel out. This is the same cancellation which we encountered in perturbation theory
in Eq.~\eqref{RenInt}, thus we conclude that the above mentioned divergent terms also cancel out in
perturbation theory if symmetries are respected during the re-summation.

To substantiate our claim we provide here a rigorous analysis of \eqref{SdQ1}. For this purpose we decompose $\delta\hat\Sigma_\alpha(\Omega,\nu)$, use the explicit form of $\Pi^{R/A}$ and take advantage of the symmetry relations \eqref{dQsym} of $R^{(i)}$.
\begin{align}
 \delta \Sigma^{11}_{\alpha}(\Omega,-\nu) &= -\frac{1}{2} \int \frac{d\omega}{2\pi} \left( 2 i g \Omega\, R^{(1)}_\alpha(\omega,\nu) + \frac{\Pi^K(\omega+\Omega)-\Pi^K(-\omega+\Omega)}{2} \left( R^{(3)}_\alpha(\omega,\nu) - R^{(4)}_\alpha(\omega,\nu) \right) \right) \notag\\
 \delta \Sigma^{12}_{\alpha}(\Omega,-\nu) &= -\frac{1}{2} \int \frac{d\omega}{2\pi} \Bigg( \frac{\Pi^K(\omega+\Omega) + \Pi^K(-\omega+\Omega)}{2} \, i R^{(1)}_\alpha(\omega,\nu)  + \frac{\Pi^K(\omega+\Omega)-\Pi^K(-\omega+\Omega)}{2} \, R^{(2)}_\alpha(\omega,\nu) -2 g \omega\, R^{(4)}_\alpha(\omega,\nu) \Bigg) \notag\\
 \delta \Sigma^{21}_{\alpha}(\Omega,-\nu) &= \delta \Sigma^{12}_{\alpha} \left(\Omega,-\nu; R^{(1)}_\alpha\rightarrow - R^{(1)}_\alpha; R^{(4)}_\alpha\rightarrow - R^{(4)}_\alpha \right) \notag\\
 \delta \Sigma^{22}_{\alpha}(\Omega,-\nu) &= \delta \Sigma^{11}_{\alpha} \left(\Omega,-\nu; R^{(1)}_\alpha\rightarrow - R^{(1)}_\alpha; R^{(4)}_\alpha\rightarrow - R^{(4)}_\alpha \right)
\label{dSigmadecomp}
\end{align}
In addition
\begin{gather}
  \Pi^{R/A}(\Omega+\omega) = \pm g\, (\Omega+\omega) -i D,\qquad \Pi^K(\Omega+\omega)= 2g (\Omega+\omega)\coth\left(\frac{\Omega+\omega}{2 T}\right) \ .
\end{gather}
At low frequencies $\Omega\ll T$ the terms containing the 'classical' contribution $\Pi^K\approx 4g T$ dominate over those 
containing $\Pi^{R/A}$. Indeed $\Pi^{R/A}$ are important only in the 'quantum' region $\Omega\gtrsim T$, then $\Pi^{R/A/K}\propto g\Omega$. Here we recall the discussion about the RG in 
Section~\ref{sec:BoseKondo}. We can disregard most of the 'quantum' domain $\Omega\gtrsim T$ because these frequencies could be integrated out in the initial RG procedure. 

Where do large contributions to the $\Omega$-integral in (\ref{SdQ1}) come from? In the region
$\Omega\sim T$ Green's functions generate a factor of $1/T^2$, the expression as a whole scales as
$g^2$ and can therefore be neglected as compared with (\ref{SdQ2}), which scales as $g$. The
remaining region to consider is $\Omega\sim \Gamma\ll T$. There, terms containing the Keldysh Green
function
\begin{align}
 G^K(\Omega) = \frac{-2 i \Gamma \tanh\frac{\Omega}{2T} }{\Omega^2 +\Gamma^2} 
\end{align}
get another order of $g$ due to the hyperbolic tangent in the small $g$ expansion: $\tanh
\Omega/(2T)\sim \Gamma/T=2g$. Neglecting $G^K$-terms we write the remaining terms of (\ref{SdQ1})
as
\begin{align}
 G^R_{0,\alpha}\delta\Sigma^{21} G^R_{0,\alpha}\delta\Sigma^{21} + G^R_{0,\alpha}\delta\Sigma^{22} G^A_{0,\alpha}\delta\Sigma^{11} + G^A_{0,\alpha}\delta\Sigma^{11} G^R_{0,\alpha}\delta\Sigma^{22} + G^A_{0,\alpha}\delta\Sigma^{12} G^A_{0,\alpha}\delta\Sigma^{12} \ .
\end{align}
In the limit $\Omega\ll T$ most prefactors in $\delta\hat\Sigma$ are linear in $\Omega$. Considering
the region $\Omega\sim\Gamma=2g T$ the linear prefactor $\Omega$ yields another order of $g$, and
therefore the corresponding terms can also be neglected. Finally, the $\Omega$-independent term
$\alpha\omega R^{(4)}$ in $\delta\Sigma^{(12)}$ only appears combined with $G^R G^R$ and $G^A G^A$,
having both poles on the same side of the real axis. The integration by residue theorem yields zero.
We conclude that all terms of \eqref{SdQ1} are of higher order in $g$ than those of (\ref{SdQ2}).
This line of reasoning still holds if a magnetic field $B$ is included in the problem provided that
$B\ll \Gamma$. For larger fields, which are however still smaller than the temperature, the
frequency in the hyperbolic tangent would essentially be replaced by $B$ and thus yield a factor
$\tanh B/(2T)\approx B/T > g$.

As a result we have confirmed the conclusion of the main text: it is justified to neglect the first
term in the action \eqref{SdQ1}, which was obtained from the expansion of the \trlog{} term around
the saddle point. The action of $\delta\hat Q$-fluctuations around the saddle point is governed by
second term, \eqref{SdQ2}, generated in the Hubbard-Stratonovich decoupling of the quartic Majorana
term with $Q$-fields. 

\section{Gauge freedom} 
\label{sec:Anonzero}

In this Appendix we explore the curious gauge freedom present in our problem. Interestingly, we find
that the saddle-point solution~\eqref{SPsol} and particularly the imaginary part of the self-energy
acquire a gauge-field dependence. As a result, the physical Green function no longer
coincides with the saddle point Green function. We find that fluctuations in turn become
important, and their role is to compensate for the effect of the introduced gauge fields.

We recapitulate the quartic term in the Majorana action \eqref{Smint_app}, before the $Q$-fields
were
introduced. 
\begin{align}
  i S_{M,int}[\eta_\alpha] &= -\frac 14 \int dtdt' \tau_3^a \, \Pi^{ab}(t,t') \,\tau_3^b \ ( \eta^a_{t,\alpha} \eta^b_{t',\alpha}) \, M_{\alpha\beta} ( \eta^a_{t,\beta} \eta^b_{t',\beta}) 
\end{align}
Due to the property $\eta_{t,\alpha}^2=0$ of Grassmann variables, adding real/complex finite entries
$A_x, A_y$ and $A_z$ on the diagonal of the matrix $M$ does not change the action. The range of
possible values of $A_\alpha$ is limited by the constraints that $M$ is invertible (or
equivalently, $\det{M}\neq0$) and its eigenvalues have to be real. We interpret the
$A_\alpha$ as gauge fields, which may be fixed by some condition.
\begin{align}
 M&= \begin{pmatrix} A_x & 1 & 1 \\ 1 & A_y & 1 \\ 1 & 1 & A_z \end{pmatrix}.
\end{align}
The redecoupling of the quartic term with the help of complex bosonic fields $Q_\alpha$ via a
Hubbard-Stratonovich transformation is not modified, the action is still of the form given in the
main text in Eq.~\eqref{SetaQ}. The only new feature consists in the diagonal non-zero entries of
the matrix $M$.
\begin{align}
 i S[\eta_\alpha,Q_\alpha] = & \int dtdt'  \left(\frac{i}{2}\eta^a_\alpha(t) \, (G_\alpha^{-1})^{ab}_{tt'} \, \eta^{b}_\alpha(t') 
 - \frac14 \tau_3^a \tau_3^b \Pi^{ab}(t,t') \, Q^{ab}_\alpha(t,t') \,M^{-1}_{\alpha\beta} Q^{ab}_\beta(t,t')\right) \ .
\end{align}
Integrating out the Majorana fermions one again obtains \eqref{SQ}
\begin{align}
 i S[Q_\alpha]= \frac 12 \text{Tr}_t \sum_\alpha^{x,y,z} \log{(\hat G_\alpha(\hat Q_\alpha))^{-1}} + \frac{1}{8}\int dt dt' \hat \Pi^{ab}(t,t') M^{-1}_{\alpha\beta}\text{Tr}\left\{ \bar\gamma^a \hat Q_\alpha(t,t')\bar\gamma^b \hat Q_\beta(t',t) \right\} ,
\end{align}
and the saddle point equations
\begin{align}
 \hat Q_{0,\alpha}(t,t')&= M_{\alpha\beta}\hat G_{0,\beta}(t,t') \notag\\
 (\hat G_{0,\alpha}^{-1})(t,t')&=  i \tau_1 \delta(t-t') \partial_{t'} - \hat
\Sigma_{0,\alpha}(t,t') \notag\\
 \hat \Sigma_{0,\alpha}^{ab}(t,t')&= -\frac 12  \text{Tr}\left\{ \bar\gamma^a \hat  \Pi(t,t')\bar\gamma^b \hat Q_{0,\alpha}(t',t) \right\} \ .
 \label{SPsolA} 
\end{align}
We emphasize that this set of equations now depends on the gauge fields $A_\alpha$, introduced
above, suggesting that there is not just one but instead a set of saddle points, characterized by
the values of $\{A_x,A_y,A_z\}$. This fact becomes obvious if we write down the self-consistent
solution in the high-temperature regime explicitly:
\begin{align}
 \Sigma^R_{0,\alpha}(\omega)&= -\frac{1}{2} M_{\alpha\beta} \int\frac{d\Omega}{2\pi} \left( \Pi^K(\omega+\Omega) G^A_\beta(\Omega) + \Pi^R(\omega+\Omega) G^K_\beta(\Omega) \right)  \notag\\
 &= -2 i  g T \left(1+\frac{A_\alpha}{2} + \mathcal{O} \left( \frac{\omega}{T} \right) \right) 
 \label{SPSigmaA} \ .
\end{align}
As the imaginary part of $\Sigma^R$ now depends on the gauge fields, we can no longer identify it as
a physically observable rate. For any finite $\{A_x,A_y,A_z\}$ equation \eqref{Sigcor} fails and
therefore the saddle point Green function and the physical Green function do not coincide, in
contrast to the result \eqref{Gphys} in the main text. In order to find the physical Green
function, we have to reconsider fluctuations around the saddle points for finite gauge fields.

To describe the fluctuations we evaluate the action (\ref{SdQ2}), that is the leading second term of $ i S_{\delta Q}$. To simplify the discussion we define
\begin{gather}
 R^{(1)}_{\alpha}(\nu)= \int_0^{\infty} \frac{d\omega}{2\pi} R^{(1)}_\alpha(\omega,\nu), \qquad R^{(i)}_{\alpha}(\nu) = \int_0^{\infty} \frac{d\omega}{2\pi} \frac{\omega}{T} R^{(i)}_\alpha(\omega,\nu) \quad \text{ for } \  i=2,3,4.  
\end{gather}
Written in the matrix form in terms of
$(R^{(1)}_\alpha,R^{(2)}_\alpha,R^{(3)}_\alpha,R^{(4)}_\alpha)$, the leading terms in the
high-temperature regime are ($i,j=1,2,3,4$)
\begin{align}
 i S_{\delta Q}^{(2)} &= 2 g M^{-1}_{\alpha\beta} \int \frac{d\nu}{2\pi} \int_0^\infty \frac{d\omega d\omega'}{(2\pi)^2} \, R^{(i)}_\alpha(\omega,\nu) \begin{pmatrix}
 2 T & 0 & 0 &  i  \omega' \\ 0 & \frac{\omega\omega'}{3 T} & 0 & 0 \\ 0 & 0 & \frac{\omega\omega'}{3 T} & 0 \\  i  \omega & 0 & 0 & -\frac{\omega\omega'}{3 T}  \end{pmatrix}_{ij} R^{(j)}_\beta(\omega',-\nu)  \\
 &= - \frac{1}{2} \int \frac{d\nu}{2\pi} \, R^{(i)}_\alpha(\nu) \, \left(D^{-1}\right)^{(ij)}_{\alpha\beta} R^{(j)}_\beta(-\nu)\ ,
 \label{SdQ2expl}
\end{align}
where
\begin{align}
 D^{(ij)}_{\alpha\beta} &= \frac{M_{\alpha\beta} }{4 g T}  \begin{pmatrix} 1 & 0 & 0 & 3 i \\ 0 & -3
& 0 & 0 \\ 0 & 0 & -3 & 0 \\ 3 i & 0 & 0 & -6  \end{pmatrix}_{ij}\quad \textrm{and} \qquad 
 \left(D^{-1}\right)^{(ij)}_{\alpha\beta} = -4 g T M^{-1}_{\alpha\beta} \begin{pmatrix} 2  & 0 & 0 & i \\ 0 & \frac{1}{3} & 0 & 0 \\ 0 & 0 & \frac{1}{3} & 0 \\ i & 0 & 0 & -\frac{1}{3}  \end{pmatrix}_{ij}\ .
\label{dQprop}
\end{align} 
Thus, we obtain for the correlator of the fluctuations
$\av{R^{(i)}_{\alpha}(\nu_1) R^{(j)}_{\beta}(\nu_2)}=D^{(ij)}_{\alpha\beta}\,2\pi\delta(\nu_1+\nu_2)$.

Knowing the propagator of fluctuations, we can compute the fluctuation-averaged Green function. As
a starting point we use the self-consistent Dyson equation, which has to be satisfied by the correct
Green function.
\begin{align}
 \left( \hat G^{-1}_{\alpha,f} - \hat \Sigma_\alpha \right) \circ \hat G_\alpha &= \mathbbm{1} \ .
\end{align}
Before the average is actually performed causality does not necessarily apply, and we have to allow
for a finite `anti-Keldysh' 22 component of the fluctuating self-energy. For the 12-component of the
Green function, which will become retarded after the averaging, we obtain the following equation
\begin{align}
 i\partial_{t} G_\alpha^{12}(t,t') = \delta(t-t') + \int dt_1 \Sigma_\alpha^{21}(t,t_1) G_\alpha^{12}(t_1,t') + 
 \int dt_1 \Sigma_\alpha^{22}(t,t_1)  G_\alpha^{22}(t_1,t') \ .
\end{align}
We have found that the propagator \eqref{dQprop} does not depend on the frequency $\omega$
corresponding to the time difference $t-t_1$. Therefore, we assume that
$\Sigma^{21}(t,t_1)=\delta(t-t_1) \Sigma^{21}(t)$ is local in time, but keep the dependence on total
time for the fluctuation average later on. We also neglect $\Sigma^{22} G^{22}$, which is of
higher order because $\Sigma^{22}\sim \Pi^{R/A} G^{R/A} +\Pi^K G^K$ is smaller than the `big'
combination $\Pi^K G^{R/A}$.
\begin{align}
 i\partial_{t} G_\alpha^{12}(t,t') =\delta(t-t') + \Sigma_\alpha^{21}(t) G^{12}(t,t')  
\end{align}
The above equation is solved by the following ansatz. The self-energy $\Sigma^{21}$ includes the
constant saddle-point contribution and fluctuations. The saddle-point part is $\Sigma^R_0$ and can
be separated. The resulting prefactor is identified with the saddle-point Green's function. 
\begin{align}
 G_\alpha^{12}(t,t') &= -i \Theta(t-t') \exp\left\{ -i\int_{t'}^t dt_1 \Sigma_\alpha^{21}(t_1) \right\}  = -i \Theta(t-t') e^{-i\Sigma_{0,\alpha}^R(t-t')} \exp\left\{ -i\int_{t'}^t dt_1 \delta\Sigma_{\alpha}^{21}(t_1) \right\} \notag \\
  &= G^R_0(t,t') \exp\left\{ -i\int_{t'}^t dt_1 \delta\Sigma_{\alpha}^{21}(t_1) \right\} 
\end{align}
We treat the $\delta\hat Q$-fluctuations using the usual Gaussian averaging procedure and use a
`self-energy correlator' for simplification.
\begin{align}
 G_\alpha^{R}(t,t') &\equiv \left< G_\alpha^{12}(t,t')\right>_{\delta Q} =  G^R_{0,\alpha}(t,t') \left<\exp\left\{-i\int^t_{t'}dt_1 \delta \Sigma_\alpha^{21}(t_1) \right\} \right>_{\delta Q} \notag\\
&= G^R_{0,\alpha}(t,t') \exp\left\{-\frac{1}{2} \int^{t}_{t'} dt_2 dt_3  \left< \delta \Sigma_\alpha^{21}(t_2) \delta \Sigma_\alpha^{21}(t_3) \right>_{\delta Q} \right\} 
\label{GdSigmaprop}
\end{align}
The 21-component of the `self-energy correlator' is decomposed into a bath correlator and
fluctuating modes according to Eq.~\eqref{dSigma}.
\begin{align*}
 \delta \Sigma^{21}_{\alpha}(t,t') &= -\frac 12  \text{Tr}\left\{ \hat \Pi(t,t')\tau_1 \delta\hat Q_{\alpha}(t',t) \right\} \\
 &= \frac{i}{2} \Pi^K(t-t') R^{(1)}_\alpha(t',t) - \frac{1}{2} \Pi^K(t-t') R^{(2)}_\alpha(t',t) - \frac{1}{2} \left( \Pi^R(t-t') + \Pi^A(t-t') \right) R^{(3)}_\alpha(t',t) \\ 
 &\qquad \qquad  - \frac{1}{2} \left( \Pi^R(t-t') - \Pi^A(t-t') \right) R^{(4)}_\alpha(t',t) \\
 &\stackrel{sym}{=} \frac{i}{2} \Pi^K(t-t') R^{(1)}_\alpha(t',t) - \frac{1}{2} \left( \Pi^R(t-t') - \Pi^A(t-t') \right) R^{(4)}_\alpha(t',t)
\end{align*}

\begin{align}
 \delta \Sigma^{21}_{\alpha}(\Omega,\nu) &= \frac{1}{2} \int \frac{d\omega}{2\pi} \Bigg( \frac{\Pi^K(\omega+\Omega) + \Pi^K(-\omega+\Omega)}{2} \, i R^{(1)}_\alpha(\omega,-\nu)  - \frac{\Pi^K(\omega+\Omega)-\Pi^K(-\omega+\Omega)}{2} \, R^{(2)}_\alpha(\omega,-\nu) - \alpha \omega\, R^{(4)}_\alpha(\omega,-\nu) \Bigg) \notag\\
 &\approx  \int_0^\infty \frac{d\omega}{2\pi} \left( 4i g T R^{(1)}_\alpha(\omega,-\nu) - 2g \omega  R^{(4)}_\alpha(\omega,-\nu) \right)  = 4 i g T R^{(1)}_\alpha(-\nu) -2g T R^{(4)}_\alpha(-\nu) \notag\\
 \Rightarrow& \quad  \delta\Sigma_\alpha^{21}(\Omega,-\nu) = 2g T \left( 2i,0,0,-1\right)\begin{pmatrix}  R^{(1)}_\alpha(\nu) \\ R^{(2)}_\alpha(\nu) \\ R^{(3)}_\alpha(\nu) \\ R^{(4)}_\alpha(\nu)  \end{pmatrix}
\end{align}

Considering the exponent of \eqref{GdSigmaprop} we find
\begin{multline}
 -\frac{1}{2}\int^{t}_{t'} dt_2 dt_3 \left< \delta \Sigma_\alpha^{21}(t_2) \delta \Sigma_\alpha^{21}(t_3) \right>_{\delta Q} = -\frac{1}{2}\int^{t}_{t'} dt_2 dt_3 \int \frac{d\nu_2}{2\pi}  \int \frac{d\nu_3}{2\pi} \left< \delta \Sigma_\alpha^{21}(\nu_2) \delta \Sigma_\alpha^{21}(\nu_3) \right>_{\delta Q}\,e^{-i\nu_2 t_2 - i \nu_3 t_3} \\
 = -\frac{(2g T)^2}{2} (t-t')\, \left(2i,0,0,-1\right)^{(i)} D^{(ij)}_{\alpha\alpha} \begin{pmatrix}  2i \\ 0\\0\\-1\end{pmatrix}^{(j)} 
 = - g T\,(t-t')\, A_\alpha \ .
\end{multline}

At the saddle point the imaginary part of the self-energy \eqref{SPSigmaA} was found to depend on the arbitrary constants $A_\alpha$. Hence it cannot correspond to the physical decay rate. We define the physical decay rates $\Gamma_\alpha$ using the physical, fluctuation-averaged $12$-component of the Green's function
\begin{align}
 i G_\alpha^R(t,t')\equiv i \left< G_\alpha^{12}(t,t')\right>_{\delta Q} = \Theta(t-t') e^{- \Gamma_\alpha  (t-t')}
\end{align}
and find 
\begin{align}
 \Gamma_\alpha&=  - \mathop{\text{Im}}\Sigma^R_{0,\alpha} - g  T  A_{\alpha} = 2g\, T.
\end{align}

In conclusion, we found that for arbitrary $A_\alpha$ the physical decay rate is not given by the
self-energy at the saddle point but rather by the decay rate $\Gamma_\alpha$ of the
fluctuation-averaged Green function, which is independent of $A_\alpha$. 

After having identified $A_\alpha$ as kind of an arbitrary gauge, we can now choose $A_\alpha=0$.
Then, the saddle-point solution coincides with the correct decay rate, and corrections due to
fluctuations cancel. In other words, the saddle point Majorana Green function coincides with the
physical Green function. Analogously this applies to the self-energy.



\bibliographystyle{elsarticle-num}
\bibliography{annals}







\end{document}